\documentclass[11pt,a4paper]{article}

\usepackage[top=2.2cm,bottom=2.2cm,left=2.3cm,right=2.3cm]{geometry}
\usepackage[T1]{fontenc}
\usepackage{microtype}
\usepackage{parskip}

\usepackage{amsmath,amssymb,amsthm}
\usepackage{bm}

\usepackage{graphicx}
\usepackage{float}
\usepackage{wrapfig}
\usepackage{caption}
\usepackage{subcaption}
\usepackage{booktabs}
\usepackage{tabularx}
\usepackage{array}
\newcolumntype{L}[1]{>{\raggedright\arraybackslash}p{#1}}
\newcolumntype{C}[1]{>{\centering\arraybackslash}p{#1}}

\usepackage[x11names,table]{xcolor}
\definecolor{guestBlue}{HTML}{1F4E79}
\definecolor{guestAccent}{HTML}{C0392B}
\definecolor{guestSoft}{HTML}{EAF2F8}
\definecolor{guestGold}{HTML}{B7950B}
\definecolor{bonnYellow}{HTML}{F5C242}

\usepackage{enumitem}

\usepackage{tcolorbox}
\tcbuselibrary{breakable,skins}

\newtcolorbox{sciencebox}[1][]{%
  breakable,
  enhanced,
  colback=guestSoft,
  colframe=guestBlue,
  boxrule=0.6pt,
  arc=2pt,
  left=8pt,right=8pt,top=6pt,bottom=6pt,
  fonttitle=\bfseries\sffamily,
  coltitle=white,
  colbacktitle=guestBlue,
  attach boxed title to top left={yshift=-2mm,xshift=4mm},
  boxed title style={sharp corners,boxrule=0pt},
  title={#1}
}

\newtcolorbox{whybox}[1][]{%
  breakable,
  enhanced,
  colback=white,
  colframe=guestGold,
  boxrule=0.7pt,
  arc=1pt,
  left=8pt,right=8pt,top=5pt,bottom=5pt,
  fonttitle=\bfseries\sffamily,
  coltitle=white,
  colbacktitle=guestGold,
  attach boxed title to top left={yshift=-2mm,xshift=4mm},
  boxed title style={sharp corners,boxrule=0pt},
  title={#1}
}

\usepackage{titlesec}

\titleformat{\section}
  {\color{guestBlue}\Large\bfseries\sffamily}{\thesection.}{0.5em}{}
\titleformat{\subsection}
  {\color{guestBlue}\large\bfseries\sffamily}{\thesubsection.}{0.5em}{}
\titleformat{\subsubsection}
  {\color{guestBlue}\normalsize\bfseries\sffamily\itshape}{\thesubsubsection.}{0.5em}{}
\titlespacing{\section}{0pt}{14pt plus 2pt}{6pt plus 1pt}
\titlespacing{\subsection}{0pt}{10pt plus 2pt}{4pt plus 1pt}
\titlespacing{\subsubsection}{0pt}{8pt plus 2pt}{2pt plus 1pt}

\renewcommand\thesubsubsection{\thesubsection.\Alph{subsubsection}}

\usepackage{fancyhdr}
\usepackage[colorlinks=true,linkcolor=guestBlue,citecolor=guestBlue,urlcolor=guestBlue]{hyperref}

\usepackage[square,numbers,sort&compress]{natbib}
\def\aap{A\&A}

\def\prd{Phys.~Rev.~D}

\newcommand{\mis}{\textsc{guest}}
\newcommand{\Mis}{\textsc{Guest}}

\newcommand{\Msun}{M_\odot}
\newcommand{\Mearth}{M_\oplus}
\newcommand{\muHz}{\ensuremath{\mu\mathrm{Hz}}}

\makeatletter
\renewcommand{\maketitle}{%
  ${}$\vspace{20pt}\begin{center}
    {\color{guestBlue}\sffamily\bfseries\LARGE \@title\par}\vspace{6pt}
    {\sffamily\large A passive satellite laser-ranging mission
       for the dark gravitational Universe\par}\vspace{10pt} 
     \includegraphics[width=0.11\textwidth]{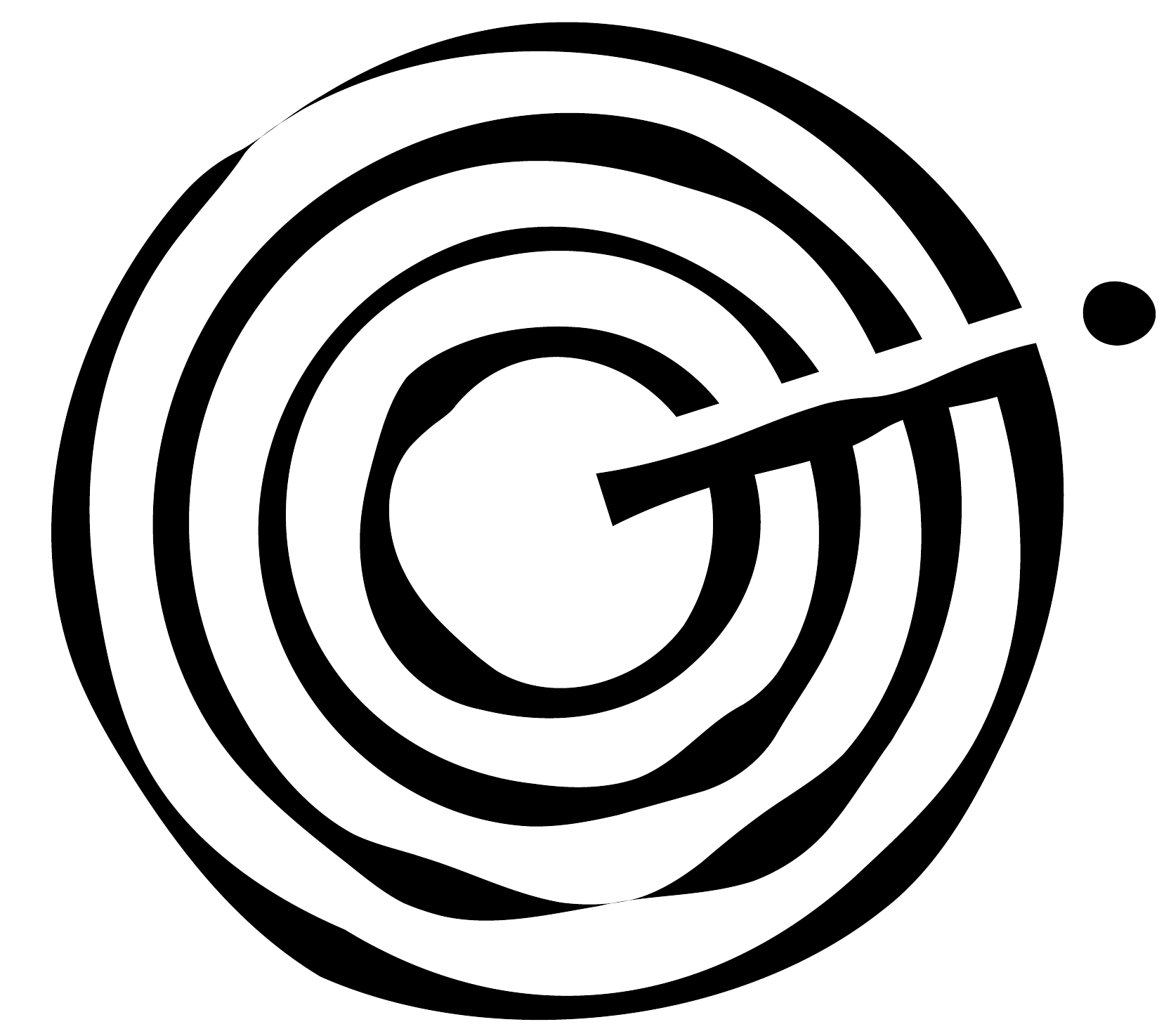}\\[8pt]
    {\small\@author}
    \vspace{2pt}
  \end{center}
  \vspace{10pt}
}
\makeatother

\usepackage{authblk}

\begin{document}

\title{GUEST: Gravitational Universe Exploration\\with Satellite Tracking}
\author[1,2$\ast$]{Diego~Blas}
\author[3]{Aur\'elien~Hees}
\author[4]{F.~Javier~Atapuerca}
\author[5]{Giada~Bargiacchi}
\author[6]{Massimo~Bassan}
\author[7,8]{Joshua N. Benabou}
\author[9]{Bruno~Bertrand}
\author[3]{Adrien~Bourgoin}
\author[10]{Clare~Burrage}
\author[5]{Nicol\`o~Burzill\`a}
\author[9,11]{Alfonso~Caldiero}
\author[5]{Roberto~Campagnola}
\author[12,13,14]{Andrea Caputo}
\author[15]{Ana~Caramete}
\author[15]{Laurentiu~Caramete}
\author[16]{Joan~Manel~Casalta~Escuer}
\author[17]{Julien~Chab\'e}
\author[15]{Gabriel~Chiritoi}
\author[18]{Marco~Cinelli}
\author[15]{Florin-Ioan~Constantin}
\author[19]{Neil J.~Cornish}
\author[17]{Cl\'ement~Courde}
\author[5]{Simone~Dell'Agnello}
\author[18]{Alessandro~Di~Marco}
\author[20]{Sebastian~Ellis}
\author[20]{Malcolm~Fairbairn}
\author[21]{Ariadna~Farr\'es}
\author[22]{Joshua~W.~Foster}
\author[12]{Silvia Gasparotto}
\author[9,11]{Marta~Goli}
\author[23]{Yann Gouttenoire}
\author[24,25]{Michael~H\"afner}
\author[26]{Antonio J. Iovino}
\author[15,27]{Maria-Catalina~Isfan}
\author[9,11]{Justin~Janquart}
\author[28,29]{Alexander~C.~Jenkins}
\author[30]{Jean-Paul Kneib}
\author[9,11]{S\'ebastien~Le~Maistre}
\author[18]{David~Lucchesi}
\author[18]{Marco~Lucente}
\author[10]{Angus~Macdonald}
\author[31,32]{Jorge~Mart\'in~Camalich}
\author[4]{Rosa~Mart\'inez~Rubiella}
\author[33,34]{Josep~J.~Masdemont}
\author[35]{Ilia Musco}
\author[36]{Toshimichi~Otsubo}
\author[1]{Crist\'obal~Padilla}
\author[37]{Fco. Rogelio Palomo Pinto}
\author[15]{Alice~Paun}
\author[38]{Alice~Perego}
\author[18]{Roberto~Peron}
\author[15,27]{Florentina-Crenguta~Pislan}
\author[15]{Florin~Adrian~Popescu}
\author[5]{Luca~Porcelli}
\author[39,34]{Nanda~Rea}
\author[31,32]{Rafael~Rebolo}
\author[31,32]{Marco~Reyes}
\author[39,34]{Ignasi~Ribas}
\author[40]{Jos\'e~C.~Rodr\'iguez}
\author[41]{Pascal~Rosenblatt}
\author[42]{Albert~Roura}
\author[9,11]{Soumen~Roy}
\author[16]{Mariano~S\'anchez~Nogales}
\author[18]{Francesco~Santoli}
\author[18]{Feliciana~Sapio}
\author[43]{Anja~Schlicht}
\author[43]{Ulrich~Schreiber}
\author[16]{Angela~Serrano}
\author[16]{Daniel~Serrano~Lombillo}
\author[44,45,46]{Alberto~Sesana}
\author[39,34]{Carlos~F.~Sopuerta}
\author[47]{Krzysztof~So\'snica}
\author[48]{Nicola~Tamanini}
\author[7,8,49]{Elisa~Todarello}
\author[12,50]{Sokratis Trifinopoulos}
\author[51]{Miguel Vanvlasselaer}
\author[5]{Dario~Vetrano}
\author[18]{Massimo~Visco}
\author[52]{Hanxi~Wang}
\author[1]{Xiao~Xue}
\author[53]{Miguel~Zumalac\'arregui}

\affil[1]{Institut de F\'isica d'Altes Energies (IFAE), The Barcelona Institute of Science and Technology, 08193 Bellaterra (Barcelona), Spain}
\affil[2]{Instituci\'o Catalana de Recerca i Estudis Avan\c{c}ats (ICREA), Passeig Llu\'is Companys 23, Barcelona, 08010, Spain}
\affil[3]{LTE, Observatoire de Paris, Universit\'e PSL, CNRS, Sorbonne Universit\'e, Universit\'e de Lille, LNE, 61 avenue de l’Observatoire 75014 Paris, France}
\affil[4]{GMV Aerospace and Defence, S.A.U., c/ de Isaac Newton 11, PTM Tres Cantos, E-28760 Madrid, Spain}
\affil[5]{Istituto Nazionale di Fisica Nucleare (INFN), Laboratori Nazionali di Frascati (LNF), Via E. Fermi 54 (già 40), 00044, Frascati, Italy}
\affil[6]{Dipartimento di Fisica e INFN, Universit\`a Tor Vergata, Roma, Italy}
\affil[7]{Theoretical Physics Group, Lawrence Berkeley National Laboratory, Berkeley, CA 94720, U.S.A.}
\affil[8]{Leinweber Institute for Theoretical Physics, University of California, Berkeley, CA 94720, U.S.A.}
\affil[9]{Royal Observatory of Belgium (ROB), Avenue Circulaire 3, B-1180 Brussels, Belgium}
\affil[10]{School of Physics and Astronomy, University of Nottingham, University Park, Nottingham NG7 2RD, UK}
\affil[11]{Universit\'e Catholique de Louvain (UCLouvain), Chemin du Cyclotron 2, 1348 Louvain-la-Neuve, Belgium}
\affil[12]{Theoretical Physics Department, CERN, Geneva, Switzerland}
\affil[13]{Dipartimento di Fisica, ``Sapienza'' Universit\`a di Roma \& Sezione INFN Roma 1, 00185 Roma, Italy}
\affil[14]{Department of Particle Physics and Astrophysics, Weizmann Institute of Science, Rehovot 7610001, Israel}
\affil[15]{Institute of Space Science - INFLPR Subsidiary (ISS), Str. Atomi\c{s}tilor 409, 077125 M\u{a}gurele, Romania}
\affil[16]{Sener Aeroespacial S.A., c/ de Severo Ochoa 4, PTM Tres Cantos, E-28760 Madrid, Spain}
\affil[17]{Observatoire de la Côte d'Azur, Universit\'e Côte d'Azur, CNRS, IRD, G\'eoazur, 2130 Route de l'Observatoire 06460 Caussols, France}
\affil[18]{IAPS - INAF Istituto di Astrofisica e Planetologia Spaziali, Via del Fosso del Cavaliere 100, 00133, Roma, Italy}
\affil[19]{eXtreme Gravity Institute, Department of Physics, Montana State University, Bozeman, Montana 59717, USA}
\affil[20]{King's College London, Strand, London, WC2R 2LS, United Kingdom}
\affil[21]{Goddard Planetary and Heliophysics Institute, University of Maryland Baltimore County, 8800 Greenbelt Rd, Greenbelt, MD 20771, USA}
\affil[22]{Department of Physics, University of Wisconsin-Madison, Madison, WI 53706, USA}
\affil[23]{Institut d’Astrophysique de Paris (IAP), CNRS, Sorbonne Universit\'e, FR-75014 Paris, France}
\affil[24]{Federal Agency for Cartography and Geodesy (BKG), Geodetic Observatory Wettzell, Bad Kötzting, Germany}
\affil[25]{Argentinean-German Geodetic Observatory (AGGO), CONICET, La Plata, Argentina}
\affil[26]{Center for Astrophysics and Space Science (CASS), New York University Abu Dhabi, PO Box 129188, Abu Dhabi, UAE}
\affil[27]{Doctoral School of Physics, Faculty of Physics, University of Bucharest, Str. Atomi\c{s}tilor 405, 077125 M\u{a}gurele, Romania}
\affil[28]{Kavli Institute for Cosmology, University of Cambridge, Madingley Road, Cambridge CB3 0HA, UK}
\affil[29]{DAMTP, University of Cambridge, Wilberforce Road, Cambridge CB3 0WA, UK}
\affil[30]{Institute of Physics, Laboratory of Astrophysics, École Polytechnique Fédérale de Lausanne (EPFL), Observatoire de Sauverny, CH-1290 Versoix, Switzerland}
\affil[31]{Instituto de Astrof\'isica de Canarias, C/ V\'ia L\'actea, s/n E38205 - La Laguna, Tenerife, Spain}
\affil[32]{Universidad de La Laguna, Departamento de Astrof\'isica, La Laguna, Tenerife, Spain}
\affil[33]{MTech \& Departament de Matemàtiques, Universitat Politècnica de Catalunya, Spain}
\affil[34]{Institut d'Estudis Espacials de Catalunya (IEEC), 08860 Castelldefels (Barcelona), Spain}
\affil[35]{Center for Astrophysics and Cosmology, University of Nova Gorica, Nova Gorica, Slovenia}
\affil[36]{Hitotsubashi University, 2-1 Naka, Kunitachi, 186-8601, Japan}
\affil[37]{Dept.\ of Electronic Engineering, Escuela T\'ecnica Superior de Ingenier\'ia, Avda.\ de los Descubrimientos s/n, Isla de la Cartuja, Universidad de Sevilla, Sevilla, Spain}
\affil[38]{Université Côte d’Azur, Observatoire de la Côte d’Azur, Laboratoire Artemis, CNRS, Bd de l’Observatoire, 06300 Nice, France}
\affil[39]{Institut de Ciències de l'Espai (ICE, CSIC), c/ de Can Magrans s/n, Campus UAB, 08193 Bellaterra, Spain}
\affil[40]{Instituto Geográfico Nacional, Red de Infraestructuras Geodésicas, Madrid, Spain}
\affil[41]{Nantes Université, Université d'Angers, Le Mans Université, Laboratoire de Planétologie et Géosciences, LPG UMR-6112, 44000 Nantes, France}
\affil[42]{German Aerospace Center (DLR), Institute of Quantum Technologies, Wilhelm-Runge-Str.~10, 89081 Ulm, Germany}
\affil[43]{Technical University of Munich, Arcisstr.\ 21, D-80333 Munich, Germany}
\affil[44]{Dipartimento di Fisica “G. Occhialini”, Università degli Studi di Milano-Bicocca, Piazza della Scienza 3, I-20126 Milano, Italy}
\affil[45]{INFN, Sezione di Milano-Bicocca, Piazza della Scienza 3, I-20126 Milano, Italy}
\affil[46]{INAF - Osservatorio Astronomico di Brera, via Brera 20, I-20121 Milano, Italy}
\affil[47]{Institute of Geodesy and Geoinformatics, Wroclaw University of Environmental and Life Sciences, Norwida 25, Wroclaw, Poland}
\affil[48]{Laboratoire des 2 infinis - Toulouse (L2IT), Université de Toulouse, CNRS/IN2P3, Toulouse, France}
\affil[49]{Universit\`a degli Studi di Torino, via P. Giuria 1, I--10125 Torino, Italy}
\affil[50]{Physik-Institut, Universität Zürich, 8057 Zürich, Switzerland}
\affil[51]{Departament de Física Quàntica i Astrofísica and Institut de Ciències del Cosmos (ICC), Universitat de Barcelona, Martí i Franquès 1, ES-08028 Barcelona, Spain}
\affil[52]{Department of Physics, Astrophysics, University of Oxford, Denys Wilkinson Building, Keble Road, Oxford, OX1 3RH, UK}
\affil[53]{Max Planck Institute for Gravitational Physics (Albert Einstein Institute), Am Mühlenberg 1, D-14476 Potsdam, Germany}
\affil[$\ast$]{Corresponding author: \texttt{dblas@ifae.es}}

 \date{\today}

\maketitle
\newpage

\begin{abstract}
\noindent
\Mis{} is a space mission concept whose central objective is the \textbf{detection of gravitational waves (GWs) in the microhertz band} --- a physics-rich frequency window that no other present or planned detector can reach at a significant level.
The concept is simple: two dense, passive spheres, covered with cube-corner
retroreflectors, deployed in \emph{highly eccentric} Earth orbits ($e \gtrsim 0.7$, period
$P \gtrsim 33$\,h), tracked continuously by the global network of satellite laser-ranging
stations over a minimum observation time of 10 years, with an expected total duration of 30 years.
The orbits themselves act as \emph{resonant detectors} of the oscillating gravitational
perturbations, with the microhertz sensitivity emerging from the selected orbital parameters. From the same data stream,
\mis{} delivers a programme of \textbf{fundamental and applied science} that cuts across
particle physics, gravitational-wave astronomy, cosmology, astrophysics, and geodesy:
the first coherent search for GWs from supermassive black-hole binaries in the
$\mu$Hz band, the exploration of primordial GW backgrounds in the unexplored energy-scale gap between
pulsar-timing arrays and LISA, a dedicated probe of ultra-light dark matter in a
parameter region untouched by any other experiment, a new way to search for ultra-light bosons, order-of-magnitude-improved tests of new gravitational interactions at astronomical ranges, and a step change in the absolute determination of $G\Mearth$ that
underpins the Global Geodetic Observing System and future navigation and
Earth-observation missions. This white paper presents the motivation, scientific
reach, and mission concept of \mis{}. 
\end{abstract}


 {\small\tableofcontents}
\newpage

\section{Why the microhertz band for gravitational waves, and why now}

Gravitational-wave (GW) astronomy in 2026 stands at a transitional point. Three main windows of observation are currently open: the
LIGO-Virgo-KAGRA (LVK) band centered at $\sim 100$\,Hz (first GW detection in 2015~\cite{LIGOScientific:2016aoc}); the pulsar timing array (PTA) band centered at nanohertz (strong evidence for GWs found in 2023~\cite{EPTA:2023fyk,Reardon:2023gzh,NANOGrav:2023gor,Xu:2023wog}); and cosmic-microwave background (CMB) observations exploring the lowest possible frequencies.  The large band of frequencies left uncovered has spurred the international community to expand upon previous capacities to probe the full extent of the gravitational-wave spectrum. This program (``adding colour to the gravitational wave sky'') promises to be just as groundbreaking as observations across the electromagnetic spectrum were in the twentieth century, as outlined in ESA's Voyage 2050 document~\citep{Voyage2050}. 

The ESA flagship mission LISA will contribute to this effort by covering the millihertz band from the late 2030s
\citep{LISA:2024hlh}, while ground-based third-generation (3G) interferometers (Einstein Telescope, Cosmic
Explorer) will further explore the audio band. Proposals to investigate the decihertz band include terrestrial atomic interferometers with large baselines~\cite{Abdalla:2024sst} or space-based laser interferometers with arms shorter than those of LISA~\cite{Kawamura:2011zz}, while GWs at frequencies above 10 kHz are currently under investigation with a variety of quantum sensor/resonator experiments~\cite{Amaral:2026bef,Aggarwal:2025noe}.

\begin{figure}[H]
  \centering
  \includegraphics[width=0.82\linewidth]{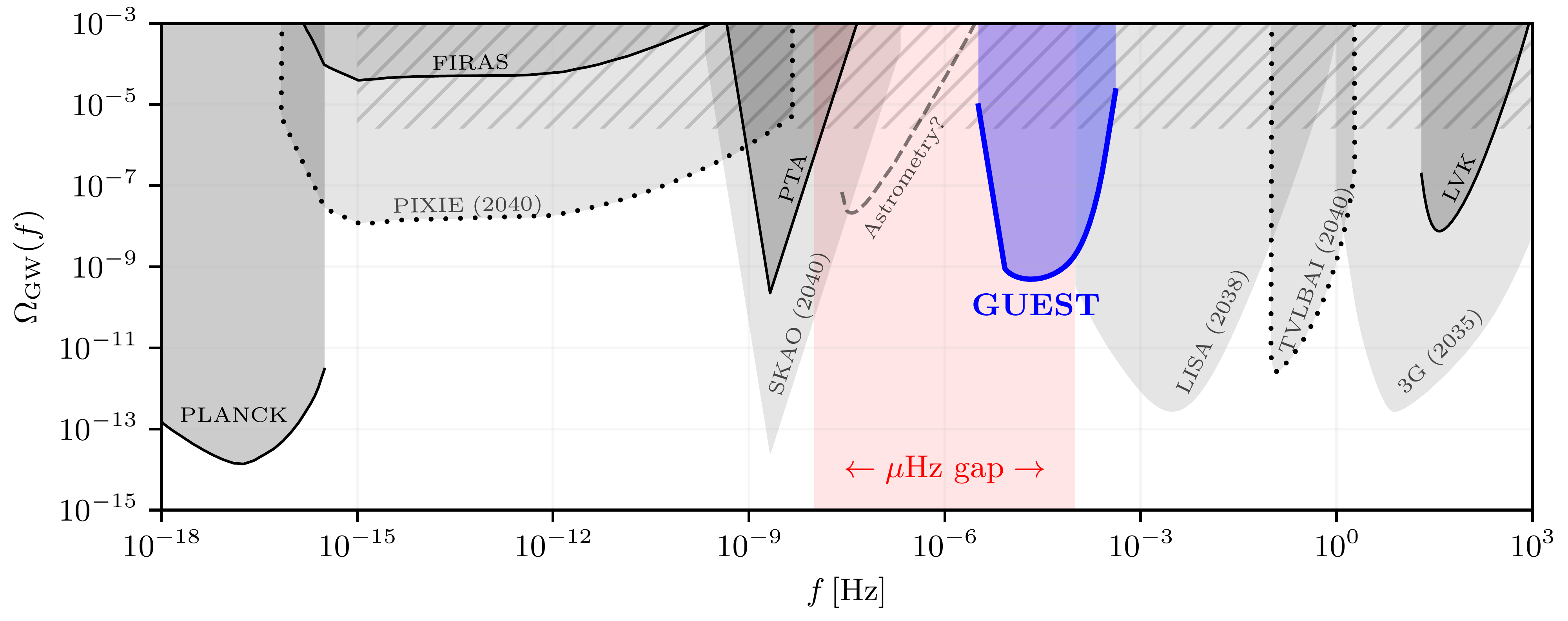}
  \caption{The gravitational-wave landscape. Current facilities (dark grey) and planned or approved
  next-generation missions (light grey) leave the $\muHz$ band essentially uncovered.
  Astrometric missions~\citep{2025arXiv250718593V} may offer complementary sensitivity, but do
  not reach astrophysically-motivated signals, and suffer from calibration issues that make robust analyses challenging. \Mis{} makes a decisive step toward closing
  this gap. TVLBAI and PIXIE lines are marked with dots and marked 2040, as they correspond to estimates, with no planned mission in the near future.}
  \label{fig:landscape}
\end{figure}

This landscape of future developments leaves the band between PTAs and LISA --- roughly four decades in frequency from $10^{-8}$ to $10^{-4}$~Hz --- \textbf{essentially unconstrained}. We refer to this
band as the \emph{microhertz gap}, see Fig.~\ref{fig:landscape}, where we have also included a \mis{} forecast.\footnote{The GW amplitude at frequency $f$ is characterised by a strain $h_0(f)$; when waves are
abundant enough to form a stochastic background, the relevant quantity is the
energy-density fraction
\begin{equation}
  \Omega_{\mathrm{GW}}(f) = \frac{1}{\rho_c}\,\frac{\mathrm{d}\rho_h}{\mathrm{d}\ln f},
  \qquad
  \rho_h = \frac{1}{32\pi G}\langle\dot h_{+}^{2}+\dot h_{\times}^{2}\rangle,
  \qquad
  \rho_c = \frac{3H^2}{8\pi G}.
\end{equation}
The \mis{} reach in both regimes is shown in Fig.~\ref{fig:sensitivity}.}  This region is expected to contain a rich variety of GW sources. In~\cite{Sesana:2019vho}, several astrophysical and fundamental signals were presented for the $\mu$Hz band, at characteristic amplitudes starting at $h_0\sim 10^{-16}$, or energy densities starting at $\Omega_{\rm GW}\sim10^{-8}$. Furthermore, the current evidence from PTA observations suggests the presence of a signal from supermassive black hole binaries 
at similar values across this band. 

\begin{tcolorbox}[ breakable,
  colframe=bonnYellow,
  colback=white,
  left=6pt, right=6pt,
width=\textwidth,
]
{\bf Why now?}
 Three factors converge to make \mis{} a mission for the near future.
\begin{itemize}\setlength\itemsep{2pt}
\item The theoretical framework that turns orbits into precision GW detectors has matured
in the past five years. The resonant response of eccentric orbits to stochastic and
monochromatic GW backgrounds is now quantitatively understood
\citep{Hui:2012yp,Blas:2021mpc,Blas:2021mqw,Foster:2025csl,Foster:2025nzf}. It is natural to work towards a mission where 
these techniques are applied to real data.
\item The global satellite laser ranging (SLR) network is simultaneously being upgraded, and the International Laser Ranging Service (ILRS) has expressed
formal support for tracking \mis{}. Several SLR stations now
demonstrate mm-level normal-point precision at altitudes of $6000$\,km; a similar result is possible for 
a few stations up to Moon distances, and up to $\sim 10^5$\,km with modest upgrades of several SLR stations.
\item A launch in the window 2030--2032 gives \mis{} a unique \emph{multi-band overlap
with LISA} (operations 2037--2041, possible extension to 2047). Sources that \mis{} sees as
quasi-monochromatic inspirals may enter the LISA band a few years later, opening a new
mode of multi-band GW astronomy.
\end{itemize}
\end{tcolorbox}

The remainder of this paper is organised as follows. Section~\ref{sec:concept} describes
the \mis{} concept, including the crucial notion of ``orbits as detectors.''
Section~\ref{sec:science} presents the science case, including, in addition to GWs, 
ultra-light dark matter, superradiance of ultra-light bosons, fifth forces, and tests
of gravity, and the geodesy return.
Section~\ref{sec:implementation} reviews the mission implementation. We close with
synergies and a summary in Sec.~\ref{sec:synergies} and Sec.~\ref{sec:summary}.

\section{The concept: orbits as detectors and their characterization}\label{sec:concept}

The \mis{} detection concept to search for GWs is closer to resonant detectors than to interferometers. In fact, rather than
measuring the \emph{instantaneous} phase shift induced by GWs, \mis{} relies on the \emph{orbits themselves} to act as a detector, which accumulates the impact from GWs over multi-year timescales by absorbing them. An analogy can be drawn with electromagnetism. An accelerating charged particle emits an electromagnetic wave and thus behaves as an emitting antenna. Conversely, an electromagnetic wave imparts acceleration to a charged particle, which then acts as a receiving antenna. The same picture applies to the gravitational sector: a binary system emits GWs, which in turn affect its orbital dynamics, making such a system the gravitational analogue of an emitting antenna. Conversely, a GW influences the dynamics of a binary system, which therefore plays the role of a receiving antenna for GWs.

\subsection{Detection concept in a nutshell}

Consider an Earth-bound satellite in motion with orbital period $P$. A GW of frequency close to the orbital frequency will have a large wavelength as compared to the orbital size (which follows from $a/P\sim v_0\ll c$, $a$ being the semi-major axis). In this approximation, the effect of the GW is to add a relative acceleration,
\begin{equation}
    \delta \ddot r^i= \frac{1}{2}\ddot h_{ij}^{\rm TT} r^j\, ,
    \label{eq:acc}
\end{equation}
where $r^i$ is the coordinate distance between the Earth and the satellite and where $h^{\rm TT}_{ij}$ is the gravitational wave metric perturbation in the transverse-traceless gauge~\citep{misner:1973fk}. Such an acceleration induces a drift of all orbital elements and, in particular, a \textbf{quadratic drift} of the true anomaly when the GW, of frequency $\omega_{\rm gw}$, resonates with the orbital motion, $P\approx n/\omega_{\rm gw}$ (with $n$ an integer)~\citep{Foster:2025csl,Foster:2025nzf}. This result is the basis of Refs.~\citep{Hui:2012yp,Blas:2021mpc,Blas:2021mqw,Foster:2025csl,Foster:2025nzf}, where the cases of binary pulsars, Earth-bound satellites, and the Moon's orbit were explored in detail. These works focused on an essentially Keplerian approach used to model the orbital dynamics. In the rest of this white paper, we go beyond this approximation and use state-of-the-art software to simulate the orbits of interest related to geodetic satellites, including most of the realistic sources of acceleration. Furthermore, we fit the GW signal simultaneously with all other parameters considered in standard Precise Orbit Determination (POD). The key message is that the analysis of~\citep{Hui:2012yp,Blas:2021mpc,Blas:2021mqw,Foster:2025csl,Foster:2025nzf} qualitatively holds as long as the resonance conditions are satisfied. 

\begin{figure}[H]
  \centering
  \begin{minipage}[t]{0.422\textwidth}
    \vspace{-9pt}
    \includegraphics[width=\textwidth]{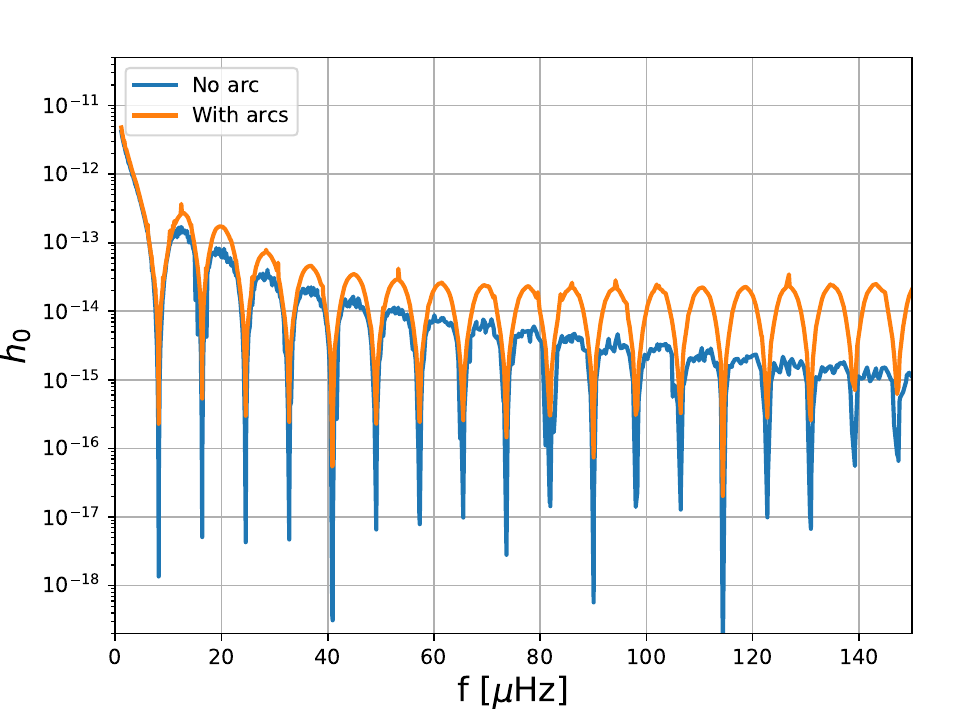}
  \end{minipage}  
    ~~~
    \begin{minipage}[t]{0.5\textwidth}
    \vspace{0pt}
    \includegraphics[width=\textwidth]{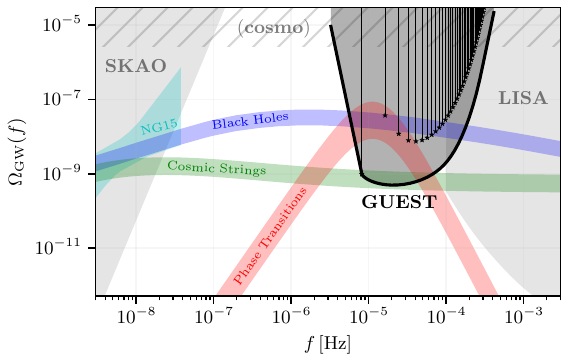}
  \end{minipage}%
\vspace{-10pt}
  \caption{\textbf{Left panel:}  Reach of \mis{} for individual monochromatic GW sources. The ``with arcs'' case corresponds to a very conservative approach to data analysis, while the ``no arc'' case is related to a best-case scenario. The curves correspond to $\mathrm{SNR}=1$.
\textbf{Right panel}: 
  Possible stochastic GW sources probed by \mis{}. Pulsar timing arrays (represented here by NANOGrav-15yr~\cite{NANOGrav:2023gor}, light blue) provide strong evidence for a stochastic GW signal from supermassive black holes~\cite{Ellis:2023owy} (blue). More speculatively, the $\mu$Hz may harbour primordial signals from the early Universe (such as cosmological phase transitions~\cite{Caprini:2015zlo}, red, or cosmic string networks~\cite{Auclair:2019wcv}, green). The black curves show the GUEST sensitivity using an ideal data analysis scheme. The continuous line is a power-law integrated (PLI) sensitivity, while the peaks correspond to the resonant sensitivity.}   
  \label{fig:sensitivity}\vspace{-0.4cm}
\end{figure}

\mis{} will leverage data from two newly deployed passive geodetic satellites.  To understand the response and sensitivity of this concept, in Fig.~\ref{fig:sensitivity}, we show the sensitivity of \mis{} to monochromatic GWs of amplitude\footnote{We emphasise that $h_0$ is the time-domain strain amplitude, and not the usual characteristic strain, which is defined in terms of a stationary power spectral density. The characteristic strain cannot be properly defined for such detectors since they are highly non-stationary.} $h_0$ and to stochastic GWs with intensity $\Omega_{\rm GW}$ (power-law integrated sensitivity~\cite{Thrane:2013oya}). \mis{}'s observational strategy is explained in the following (Sec.~\ref{sec:mission}).   

\subsection{Mission concept and data acquisition}\label{sec:mission}

The orbital parameters for the \mis{} spacecraft (S/C) follow from an optimization based on four requirements: \emph{(i) maximum eccentricity} to accelerate the quadratic growth and enhance the effect, and to excite higher resonances~\citep{Blas:2021mpc,Blas:2021mqw,Foster:2025csl,Foster:2025nzf}  ; \emph{(ii) long period} to reach the frequencies in the $\mu$Hz gap ; \emph{(iii) visibility and SLR reach} of the S/C from ground station to optimize data acquisition and \emph{(iv) minimisation of possible systematic accelerations} that may complicate the extraction of the GW signal.

Our observable is the two-way time of flight of short laser
pulses between a satellite laser ranging (SLR) station on Earth and satellites in the selected orbits. 
The satellites are passive spheres covered by
cube-corner retroreflectors (CCRs), whose centre-of-mass positions can be determined to sub-cm precision from the ranging data. To this end, the measurements are averaged into \textbf{normal points} (NPs), which are standardized observables derived from data collected over intervals of a few minutes, thereby mitigating different short-term disturbances. This SLR technique has been successfully developed since the 1970s, with the \href{https://lageos.gsfc.nasa.gov/}{LAGEOS}  and
\href{https://ilrs.gsfc.nasa.gov/missions/satellite_missions/current_missions/lars_general.html}{LARES} satellites being two of the leading missions. The final data stream is hence composed of a time series of NPs collected from different stations. The \href{https://ilrs.gsfc.nasa.gov/}{International Laser Ranging Service} (ILRS) has the mission of coordinating the data products and targets from several stations worldwide. 

\begin{figure}[H]
  \centering
  \includegraphics[width=0.8\linewidth]{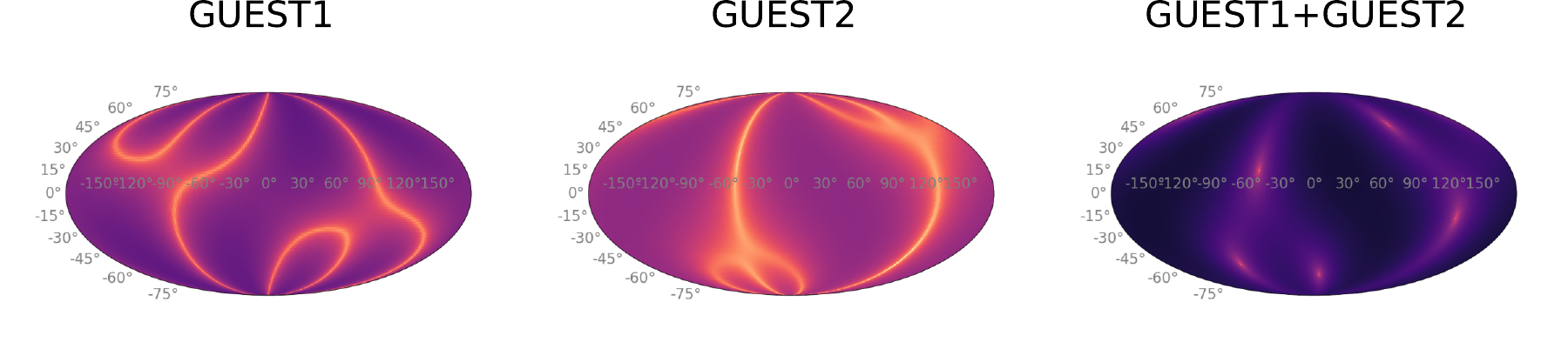}\vspace{-0.3cm}
  \caption{Sensitivity to monochromatic GWs as a function of sky location for one month of data, for each of the two \mis{} spacecraft separately and for their combination. The colour code is such that darker means stronger bounds, while the global scale is not relevant. The
  single-spacecraft response is anisotropic; the joint response, and the natural drift of
  the orbital planes over the mission, restore near-uniform full-sky coverage. }
  \label{fig:skymap}\vspace{-0.5cm}
\end{figure}

The choice of possible orbits for \mis{} is also driven by the satellites' size and mass, as this directly impacts their optical visibility, orbital dynamics and tracking performances.  For concreteness, we will frame \mis{} within the envelope of \href{https://www.cosmos.esa.int/web/call-for-missions-2025}{F-class ESA} missions or \href{https://explorers.larc.nasa.gov/APSMEX26/SMEX/}{SMEX NASA} missions. 
A particularly relevant aspect for this consideration is that the technology readiness level (TRL) of \mis{} is very high, such that a fast and low-cost concept is realistic. With these boundary conditions in mind, we propose a baseline scenario involving the deployment of two passive spheres in the orbits specified in Table~\ref{tab:orbits}. An orbital period of 33.8\,h provides access to frequencies as low as $\sim\!10^{-5}$~Hz for the lowest harmonics, well into the $\muHz$ band. The choice of inclination and right ascension of the ascending node (RAAN) may appear unusual; however, the rationale behind the specific parameters listed in Table~\ref{tab:orbits} is twofold. First, they are chosen to make the satellites \textit{naturally reenter} after 30 years. This choice was made to comply with the debris mitigation requirements of ESA (see \href{https://sdup.esoc.esa.int/documents/download/ESSB-ST-U-007_Issue_1_Revision_1_23_October_2025.pdf}{ESSB-ST-U-007} and \href{https://esamultimedia.esa.int/docs/spacesafety/ESA_Space_Debris_Mitigation_Compliance_Verification_Guidelines.pdf}{ESSB-HB-U-002}). Second, the relative geometry of the two orbits allows one to decorrelate some of the effects not related to the science case and generate a more complete sky coverage of the signals, cf. Fig.~\ref{fig:skymap}.
\begin{table}[H]
\centering\small
\begin{tabular}{@{}lccccccc@{}}
\toprule
Orbit (at $t_0$) & $h_\mathrm{perigee}$ & $h_\mathrm{apogee}$ & AOP & RAAN & $e_0$ & $i$ & $P$ \\
\midrule
\textbf{GUEST-1} & 6\,926 km & 86\,700 km & $270^\circ$ & $0^\circ$   & 0.75 & $75^\circ$ & 33.8 h \\
\textbf{GUEST-2} & 6\,926 km & 86\,700 km & $270^\circ$ & $220^\circ$ & 0.75 & $50^\circ$ & 33.8 h \\
\bottomrule
\end{tabular}
\caption{Nominal \mis{} orbits at the reference epoch $t_0$=2027-01-01. The two orbits
share the same shape but differ in plane orientation, maximising decorrelation of the GW
signal and complementary sky-coverage. Natural re-entry occurs after $\sim$29--32 years.}
\label{tab:orbits}\vspace{-.3cm}
\end{table}

The estimates in this document assume an observation campaign that tracks the orbits in Table~\ref{tab:orbits} at a rate of 10~NPs per orbit, for 10 years at a precision of 10 cm. This is conservative in a few senses: one is that the mission is assumed to last for an extended period of 30 years; furthermore, better precision (up to cm) and more NPs are realistic possibilities that would generate better results. 

In all analyses performed in this paper, we considered two scenarios: $(i)$ a conservative scenario where the POD is performed using 7-day arcs  (meaning that the initial conditions of each S/C are re-estimated every 7 days), and $(ii)$ an optimistic scenario where we analyse the data using a single 10-year-long arc. The inclusion of arcs is currently needed in geodesy analysis to handle orbit mismodeling; see e.g. \citep{Strugarek2026arcs}, but it is highly unfavorable for long-term signals. We believe this approach is overly pessimistic (or conservative) and can be improved. On the other hand, a single global solution for the full mission lifespan is likely overly optimistic, but it gives an order-of-magnitude estimate of a best-case sensitivity if the orbital modeling is improved. The differences between both methods can be seen in the left panel of Fig.~\ref{fig:sensitivity}.

\subsection{End of mission}\label{sec:end}

As mentioned briefly above, the selected orbits are such that the S/C \textit{naturally reenter} the atmosphere after 30 years. This choice is a compromise that complies with the
debris mitigation requirements of ESA (see \href{https://sdup.esoc.esa.int/documents/download/ESSB-ST-U-007_Issue_1_Revision_1_23_October_2025.pdf}{ESSB-ST-U-007} and \href{https://esamultimedia.esa.int/docs/spacesafety/ESA_Space_Debris_Mitigation_Compliance_Verification_Guidelines.pdf}{ESSB-HB-U-002}) while delivering interesting physics on a decade-long timescale. The nominal phase of the mission, which corresponds to the time in which the first scientific milestones will be achieved, is 10 years. An extended phase of 20 further years will allow \mis{} to consolidate its scientific return and increase the number of possible GW detections at very low cost. 

\begin{figure}[h]
  \centering\vspace{-.3cm}
  \includegraphics[width=0.8\linewidth]{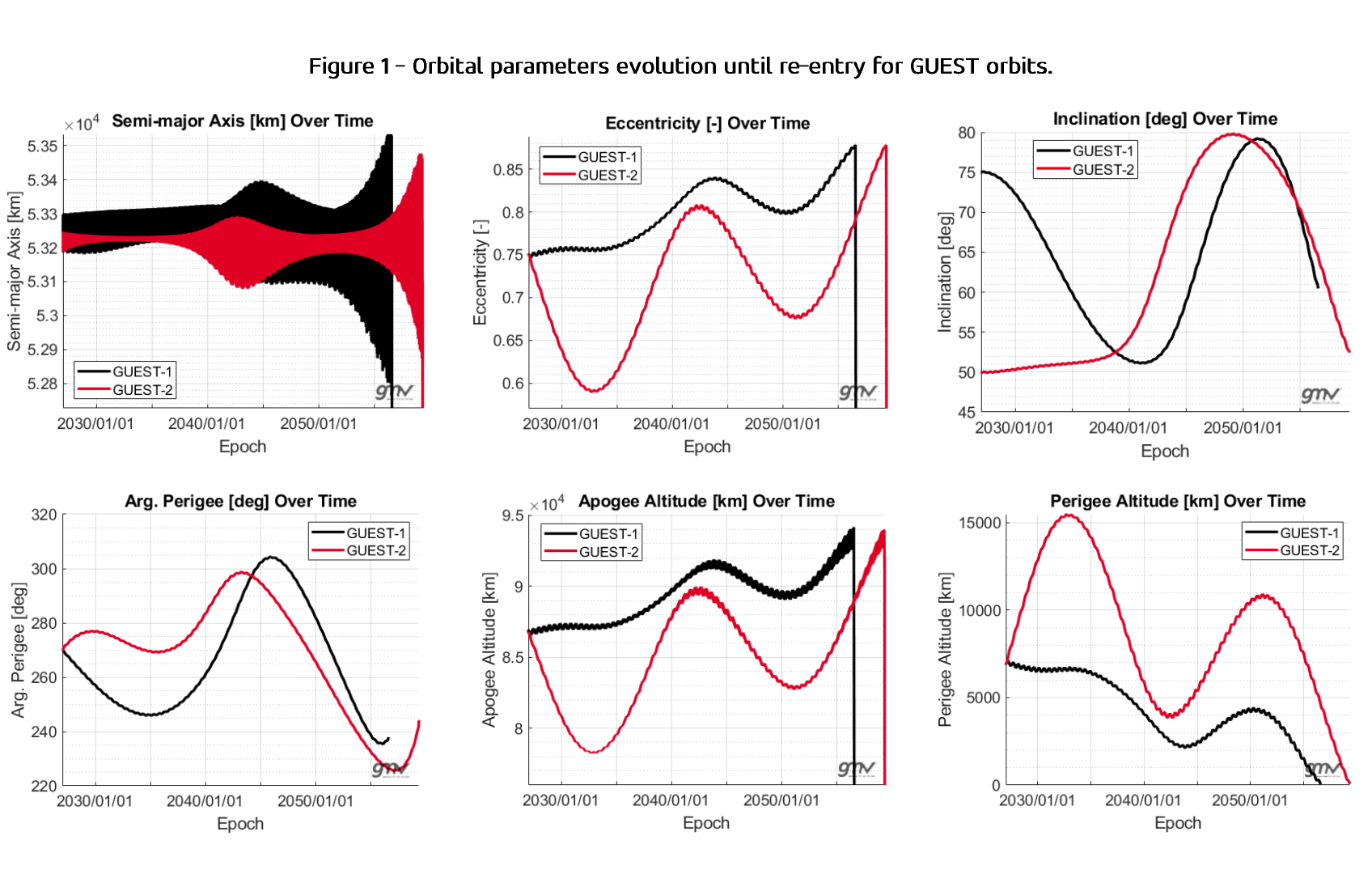}
  \caption{Evolution of the two \mis{} orbital elements over the full 30-year mission. The large variations are
  driven by lunisolar (Lidov--Kozai) dynamics. The natural increase of eccentricity
  produces atmospheric re-entry at mission end, without active disposal.}
  \label{fig:orbits}
\end{figure}

After 30 years, the eccentricity of both orbits grows until the perigee touches the atmosphere, and soon after the satellites reenter, cf.~Fig.~\ref{fig:orbits}.  The S/C design is such that they do not disintegrate when re-entering the atmosphere and remain intact as a solid block, as confirmed by simulations with the DRAMA (Debris Risk Assessment and Mitigation Analysis) software (available at \url{https://sdup.esoc.esa.int/}). The probability of Expected Casualty Risk is $E_c \sim 2$--$5\times 10^{-5}$, below the $10^{-4}$ threshold imposed by ESA requirements.   

\vspace{-.3cm}
\section{GUEST Science}\label{sec:science}

Having established the properties of the orbits and the proposed observing strategy, we can now more concretely describe the science that will be pursued. Our aim is not to provide a detailed motivation for each scientific case, but to showcase some clear directions where \mis{} can have a relevant impact.

\subsection{Gravitational-wave science in the microhertz band}

\subsubsection{Supermassive black-hole binaries: the galactic engine}
\label{sec:smbhbs}

Supermassive black-hole binaries (SMBHBs) play a pivotal role in galaxy evolution. After a galactic merger, the two central black holes sink by dynamical friction and eventually merge by a combination of stellar scattering, gas-disc torques, and --- at sub-parsec separations --- GW emission. The GW-driven inspiral takes SMBHBs with chirp masses $10^6$--$10^9\,\Msun$ \emph{directly through the microhertz band} before they merge, with the emission peaking between $\muHz$ and $\mathrm{mHz}$ depending on the mass of the binary. 

The recent PTA results~\citep{NANOGrav:2023gor} provide strong evidence for a stochastic signal at nanohertz frequencies that is compatible with SMBHB emission. Confirming this origin --- or ruling it out in favour of a primordial interpretation --- requires observations at other frequencies. \Mis{} naturally fulfills this role: as the right panel of Fig.~\ref{fig:sensitivity} shows, the predicted SMBHB spectrum (blue band) extends directly into the \mis{} sensitivity band, and the two probes together provide the lever arm needed
to discriminate among population-synthesis scenarios.

\begin{sciencebox}[Science goal SGA1 --- stochastic SMBHB background]
\Mis{} will be sensitive to the expected stochastic background of GWs from a population of supermassive black-hole binaries, currently uncharacterised at $\muHz$ frequencies.
\end{sciencebox}

\begin{figure}[H]
  \centering
  \begin{minipage}[t]{0.9\textwidth}
    \vspace{-9pt}~~~~~~~~~~~~~~~~~~~~~~~
    \includegraphics[width=0.5\textwidth]{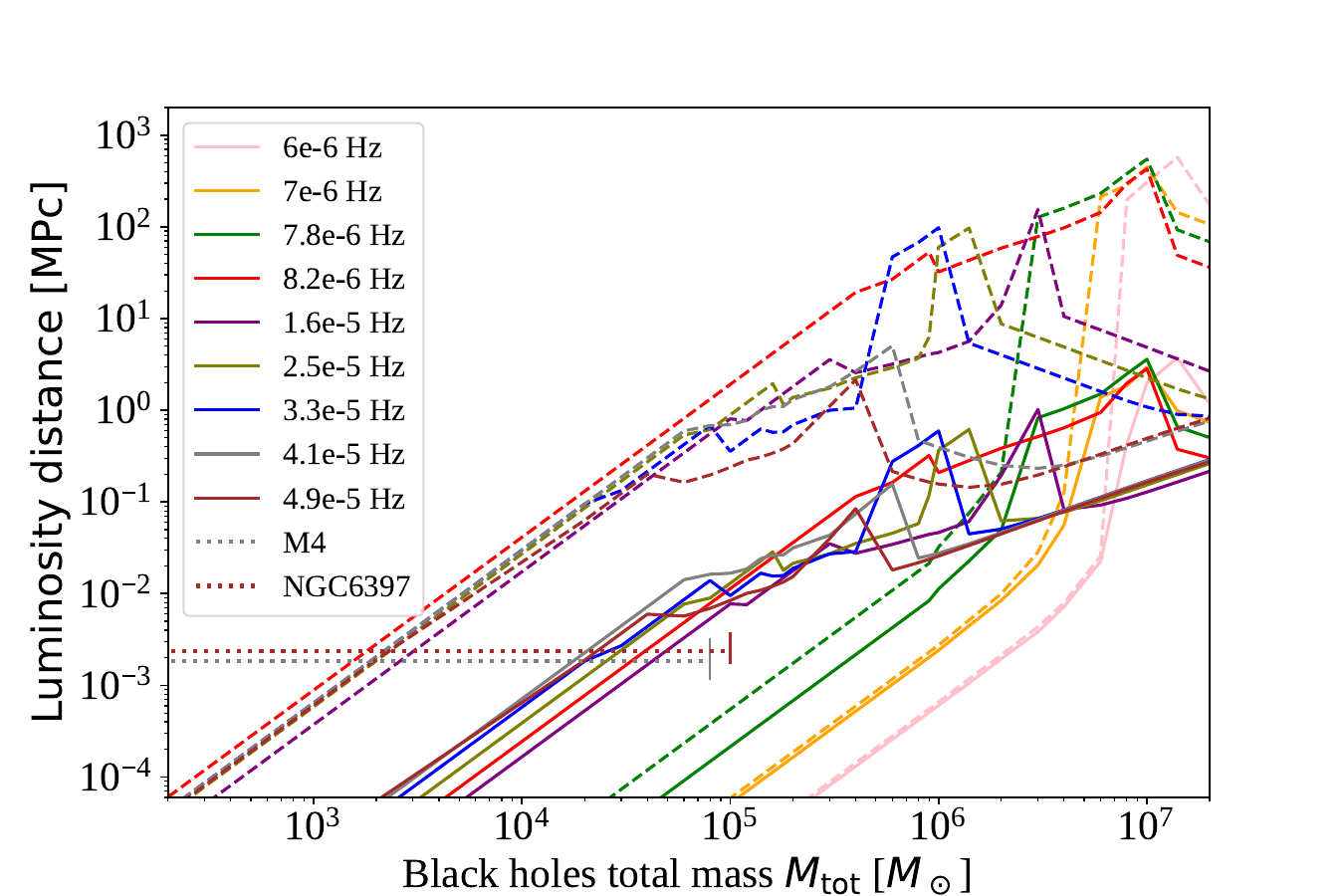}
  \end{minipage}  \\
    \begin{minipage}[t]{0.58\textwidth}
    \vspace{8pt}\hspace{-1cm}
    \includegraphics[width=1.1\textwidth]{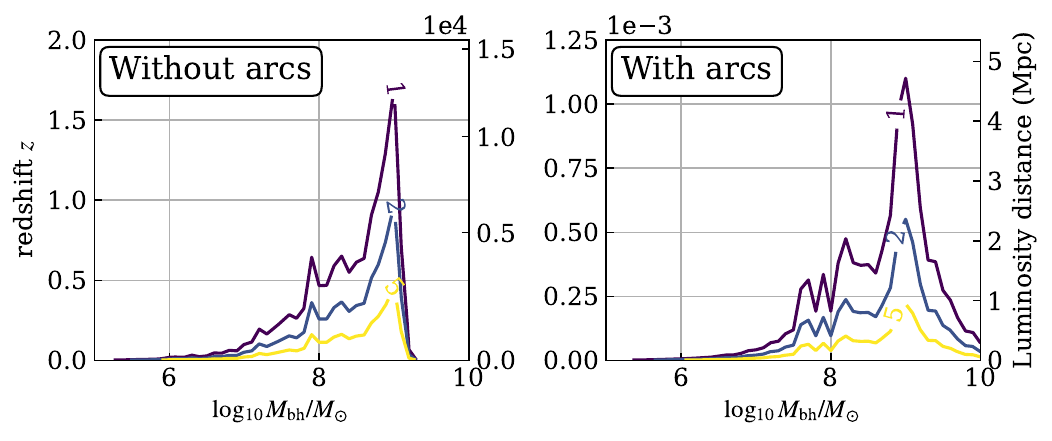}
  \end{minipage}%
  \caption{\textbf{Upper panel:}  \Mis{} distance reach for slowly evolving SMBHB sources. Colour indicates the
  binary frequency at the start of the mission; solid and dashed lines indicate conservative and optimistic data-analysis scenarios (7-day arcs and 10-year arcs, respectively).
\textbf{Lower panels}: \Mis{} distance reach for merging SMBHBs. The ``with arcs'' curve corresponds
  to the conservative data-analysis scenario; coloured curves are SNR=1, 2, 5 contours. }
  \label{fig:smbh_reach}\vspace{-0.2cm}
\end{figure}

For loud sources that stand out above the confusion background, \mis{} offers a second observational mode: \emph{individual source resolution}. If the binary is far from merger, the GW is quasi-monochromatic over the mission duration, and the response of the \mis{} orbits is strongly resonant at harmonics of the orbital period, cf. Fig.~\ref{fig:sensitivity} left panel. This leads to the distance reach shown in the left panel of Fig.~\ref{fig:smbh_reach}, which provides an effective sensitivity over the total detection time. In this graph, each frequency interval $\Delta f$ of Fig.~\ref{fig:sensitivity} is converted into a time interval $\Delta t(f;\Delta f)$ on which the system generates a resonant response, by considering the inverse of the frequency evolution $f(t)$ of a binary system of a given total mass.
For intermediate-mass black hole binaries, with a total mass $10^2-10^5\, \Msun$, the sources at frequencies on the order of $10^{-5}$ Hz are quasi-monochromatic. Fig.~\ref{fig:smbh_reach} shows that \mis{} would be able to detect any intermediate mass binary black hole (IMBH) system within our closest globular
clusters such as, for example, M4 or NGC6397. Detecting binary IMBHs would fill one of the biggest gaps in black hole astrophysics: the range between stellar-mass black holes and
supermassive black holes. 

For SMBHBs, whose frequency evolution is significant over \mis{}’s 10-year mission, a succession of resonance peaks rings up as the orbit evolves: the \mis{} response acts as a natural spectrograph in time. For the most massive binaries, \mis{}’s frequency spectrum is fully swept in a matter of months, or even hours, with a merger frequency above \mis{}’s frequency range for the cases addressed in the left panel of Fig.~\ref{fig:smbh_reach}. Finally, the merger's frequency may fall within the \mis{}'s frequency range. Fig.~\ref{fig:smbh_reach}, middle and right panels, obtained using dedicated waveforms for this regime, indicates that \mis{} is sensitive to
supermassive binaries up to $10^9\,\Msun$ up to $z\sim 1$. These are among the most massive merging systems that may ever be observed, and a measurement of their merger dynamics tests strong-field general relativity in a regime inaccessible at kilohertz frequencies.
More details on these and other estimates for individual sources will appear in future work.

\begin{sciencebox}[Science goal SGA2 --- individual SMBHB/EMRI sources]
\Mis{} will achieve first-ever sensitivity to \emph{individual} sources of GWs in the
$\muHz$ band, including quasi-monochromatic binaries of massive black holes, chirping
mergers up to $\sim 10^9\,\Msun$, and intermediate-mass black-hole binaries in
nearby globular clusters. Some of these sources may later enter the LISA band,
enabling multi-band GW astronomy.
\end{sciencebox}

Besides the previous possible sources, it is important to recall that the history of astronomy is punctuated by the discovery of unexpected sources every time a new frequency band has been opened. Among the speculative --- and exciting --- possibilities
specific to the microhertz band, we highlight: \textit{i)} A possible 10 $M_\odot$ companion of Sgr~A$^*$ would emit in \mis{}'s band
\citep{gourgoulhon:2019aa,Sesana:2019vho};  \textit{ii)} extreme mass-ratio inspirals (EMRIs), which carry detailed information about the
background spacetime of the host SMBH~\citep{Sesana:2019vho}.

\subsubsection{Primordial gravitational waves: the heart of the Big Bang}
\label{sec:primordial}

Gravitational waves propagate freely from the Universe's earliest moments, 
carrying imprints of processes associated with the extreme energy scales of the hot Big Bang. Three categories of early-Universe sources generate GW backgrounds of particular
interest in the \mis{} band~\citep{LISA:2024hlh,Caprini:2018mtu}: 

\textbf{(i) Inflation.} The standard inflationary paradigm~\citep{Caprini:2018mtu} predicts a broadband stochastic background from the amplification of
vacuum fluctuations. From a \mis{} perspective, non-standard inflationary scenarios --- axion inflation, features in the potential, transitions to
non-slow-roll phases, or the tail of a scalar power spectrum producing primordial black
holes --- generate features with power in the $\mu$Hz band.

\textbf{(ii) First-order phase transitions.} A phase transition at a temperature
$T_\star$ produces a GW spectrum peaked at $f_\mathrm{peak}\sim 19\,\mu\text{Hz} \times \frac{T_*}{100\,\text{GeV}}\,\frac{\beta/H_*}{v_w}\left(\frac{g_*}{106.75}\right)^{1/6}$. The parameters $\beta$ and $v_w$ are related to the nature of the transition, while $H_*$ is the rate of cosmological acceleration at the transition epoch~\citep{Caprini:2018mtu}. As a result, models of the early Universe where such a transition happens around 1--$100\,$GeV may leave an observable imprint in the \mis{} band. 
As a result, a \mis{} detection would constrain new physics at a range of energies connected to the electroweak scale.

\textbf{(iii) Topological defects.} Networks of cosmic strings --- generic relics of
gauge-symmetry breaking at high energy --- emit a broad, roughly scale-invariant GW background
whose amplitude is set by the string tension $G\mu$~\cite{Auclair:2019wcv}.
The extremely broadband nature of this signal means that, if it exists, it is guaranteed to pass through the \mis{} frequency band.

All of these potential cosmological sources (as well as a plethora of models beyond those mentioned here) encode vital information about the physics of the early Universe in their spectral shapes and amplitudes.
Distinguishing between these signals and decoding this information can therefore be extremely challenging if one's data is confined to a narrow frequency band, as evidenced by the large number of different models that have been proposed as sources of the nHz background probed by PTAs~\cite{NANOGrav:2023hvm}.
Exploring the early Universe therefore requires a broadband observational effort across the GW spectrum.
\Mis{} is poised to play an important role in this effort, probing unique regions of parameter space that are inaccessible to other experiments (e.g., particular energy scales associated with phase transitions or with decaying topological defects), as well as confirming or ruling out signal models that are favoured by other experiments, such as PTAs or LISA.

\begin{sciencebox}[Science goal SGA3 --- cosmological GW backgrounds]
\Mis{} will probe new regimes of high-energy phenomena in the early Universe, including
non-standard inflation, first-order phase transitions, and the dynamics of topological defects.
These signals may be
directly connected to new fundamental particles, forces, and symmetries.
\end{sciencebox}

\begin{whybox}[The synergy with LISA: inspiralling events]
LISA will cover the frequency band of $2\times 10^{-5}$--$1$~Hz from the mid-2030s at different levels of accuracy~\cite{LISA:2024hlh}.
\Mis{}, operating between 2034 and $\sim 2066$ and with a frequency window of $10^{-7}$--$10^{-4}$~Hz will be strongly complementary to LISA. A chirping SMBHB that \mis{} sees at the start of its mission
may cross into the LISA band a few years later, providing simultaneous multi-band
coverage and improved sky localisation.
The possibility of having certified LISA sources will also allow for archival searches and calibration of \mis{}.
\end{whybox}

\subsection{Dark matter}\label{sec:dm}

Dark matter (DM) composes 85\% of the gravitating matter in the Universe, but its fundamental nature is unknown. If it is a manifestation of a new fundamental particle, its mass can be as low as $m_\mathrm{DM}\!\sim\!10^{-22}$~eV\!/c$^{2}$ and as high as the largest conceivable energy for particle physics~\citep{EUreport,APPEC}. Another possibility is that DM is made of compact objects, such as primordial black holes. \Mis{} can be key to exploring both possibilities. 

\subsubsection{Coherent field oscillations: direct detection in the Solar System}

For particle DM of mass below $m_\mathrm{DM}\approx {\rm  eV/c^2}$, DM in the Milky Way behaves as a \emph{classical field}, producing coherent oscillations that may translate into oscillating forces among gravitating bodies similar to that in Eq.~\eqref{eq:acc}~\citep{Foster:2025nzf}.
The force will oscillate at a frequency,
\begin{equation}
  f_\mathrm{DM} = \frac{m_\mathrm{DM}c^{2}}{2\pi\hbar }
  \approx 2\times 10^{-6}~\mathrm{Hz}
  \left(\frac{m_\mathrm{DM}}{10^{-20}~\mathrm{eV} c^{2}}\right),
  \label{eq:ULDMfreq}
\end{equation}
or at double this frequency depending on the nature of the DM coupling sourcing the oscillations.

Figure~\ref{fig:uldm}, left panel, shows the \mis{} reach in the $(m_\mathrm{DM},
\beta\rho_\mathrm{DM}/\rho_\odot)$ plane, where $\rho_\mathrm{DM}$ is the local DM density, $\rho_\odot = 0.4~\mathrm{GeV}/\mathrm{cm}^3$ and $\beta$ parameterises an effective quadratic coupling to
baryonic matter. The case $\beta\!=\!1$ corresponds to a universal gravitational coupling, while higher $\beta$ corresponds to other model-dependent interactions. For models in which the ULDM couples more strongly to ordinary matter (e.g.\ dilaton
models, B$-$L gauge bosons, scalar-mediated fifth forces), the sensitivity improves
correspondingly, and \mis{} becomes a competitive discovery probe. 

\begin{sciencebox}[Science goal SGB1 --- ULDM via coherent oscillations]
\Mis{} will extend the direct search for ultra-light dark matter in the Solar System over
a mass range and coupling space currently unconstrained by any experimental or astrophysical probe.
\end{sciencebox}

\subsubsection{Superradiant boson clouds}

Another way to detect a new ultra-light boson (whether or not it is responsible for dark matter) is based on the phenomenon of black hole superradiance. When a bosonic field's Compton wavelength is
comparable to the gravitational radius of a rotating black hole, a superradiant
instability~\citep{Brito:2015oca} may extract the rotational energy of the black hole,
building up a macroscopic, quasi-stationary \emph{boson cloud} bound to the black hole.
The cloud subsequently radiates gravitationally at a well-defined frequency,
\begin{equation}
  f_\mathrm{GW} \approx \frac{m_\mathrm{b} c^{2}}{\pi\hbar}
  \simeq 5\times 10^{-6}~\mathrm{Hz}
  \left(\frac{m_\mathrm{b}}{10^{-20}~\mathrm{eV}/c^{2}}\right),
\end{equation}
corresponding to $2f_\mathrm{DM}$ of Eq.~\eqref{eq:ULDMfreq}.
 For bosons in the mass range
$10^{-20}$--$10^{-18}~\mathrm{eV}/c^2$, the GW emission falls in the \mis{}
band, and the corresponding black holes may be \emph{nearby supermassive} black holes.

\begin{figure}[H]\vspace{-20pt}
  \centering
  \begin{minipage}[t]{0.48\textwidth}
    \includegraphics[width=0.9\textwidth]{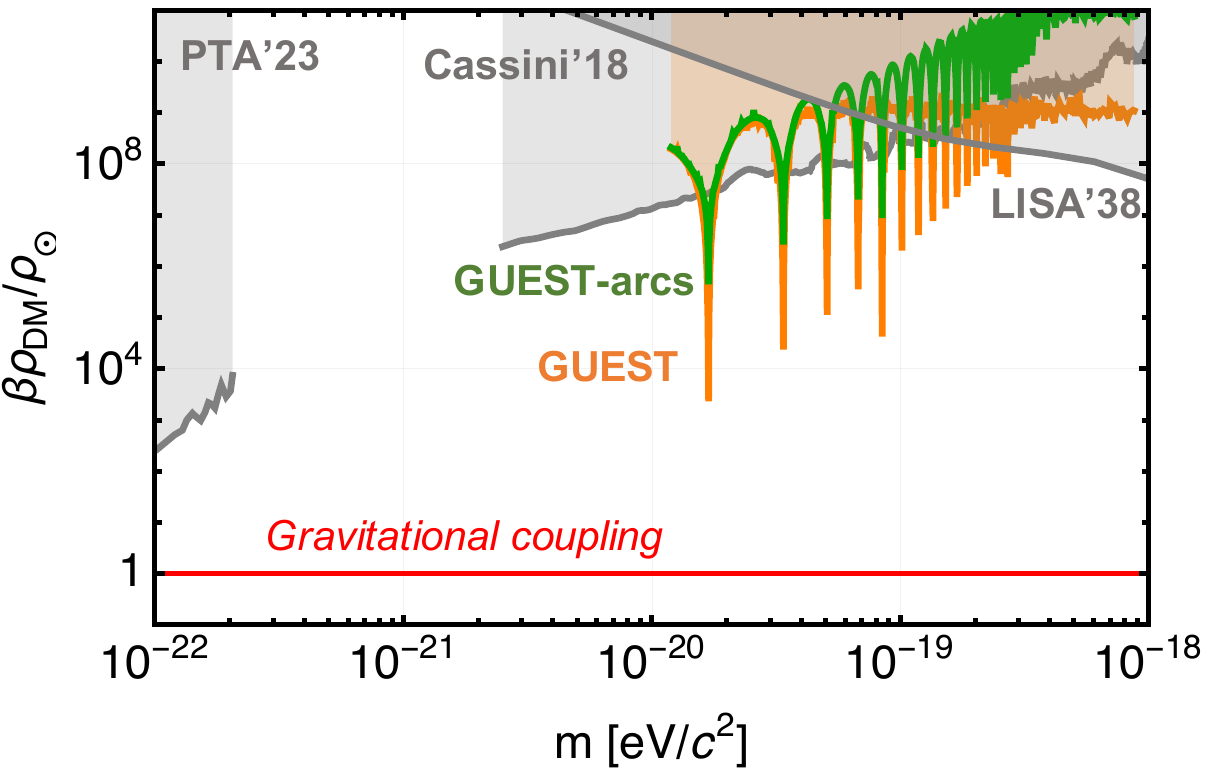}
  \end{minipage}  
    \begin{minipage}[t]{0.48\textwidth}
    \includegraphics[width=0.85\textwidth]{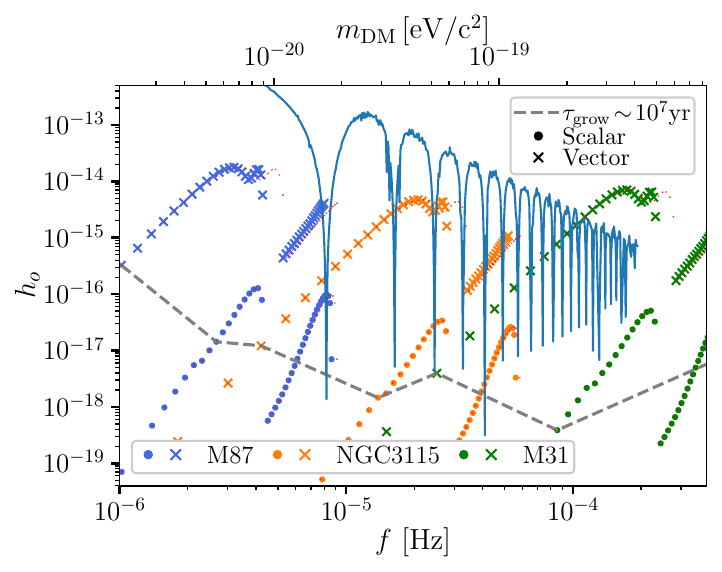}
  \end{minipage}%
\vspace{-6pt}
  \caption{\textbf{Left panel:}  \Mis{} reach for ULDM detection via coherent gravitational oscillations, in the
  $(m_\mathrm{DM},\,\beta\rho_\mathrm{DM}/\rho_\odot)$ plane. $\beta\!=\!1$ corresponds
  to the universal (unavoidable) gravitational coupling. Bounds from PTAs, Cassini, and expected LISA constraints are shown for comparison. The unshaded region is open
  parameter space.
\textbf{Right panel}: Predicted strain from scalar and vector boson clouds around three nearby
  supermassive black holes. The dashed grey line marks a cloud growth time of
  $10^7$~yr, comparable to a typical AGN lifetime.}   
  \label{fig:uldm}\vspace{-0.6cm}
\end{figure}

Figure~\ref{fig:uldm}, right panel, shows the predicted strain of the emitted GWs for scalar and
vector clouds around the black holes M87, NGC~3115, and M31. The striking feature is that \emph{for every
target, vector clouds produce signals stronger than scalar clouds}, and \emph{both}
signal classes fall within the \mis{} sensitivity region. Notice that over the mission duration, the
signals are effectively monochromatic.

\begin{sciencebox}[Science goal SGB2 --- superradiant boson clouds]
\Mis{} will enable searches for new ultra-light bosons via the gravitational-wave
emission from superradiant clouds around nearby supermassive black holes, in the mass
range $10^{-20} \lesssim m_\mathrm{b}/(\mathrm{eV}/c^{2})\lesssim 10^{-18}$, a regime
inaccessible to any other GW detector.
\end{sciencebox}

\subsubsection{Imprints from primordial black holes}

\begin{wrapfigure}{R}{0.42\textwidth}
\vspace{-28pt}
  \centering
  \includegraphics[width=0.41\textwidth]{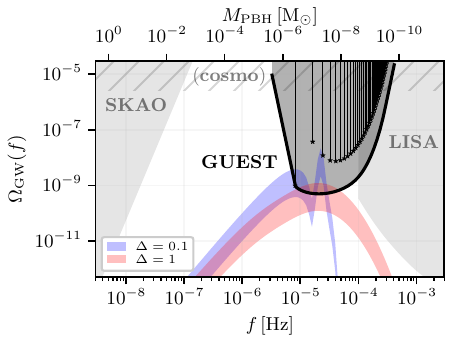}\vspace{-8pt}
  \caption{PLI sensitivity curve of \mis{} vs. the spectra for benchmark scenarios of PBHs. $\Delta$ is the width of the initial overdensity \cite{Blas:2026xws}.}
  \label{fig:PBH}
\vspace{-14pt}
\end{wrapfigure}

Part or all of DM may be in the form of primordial
black holes (PBH)~\cite{Carr:2026hot}. In the standard
formation scenario, PBHs arise from the collapse of large-amplitude curvature
perturbations at horizon re-entry, and the same perturbations unavoidably
source a stochastic background of scalar-induced
gravitational waves (SIGWs) peaked at
$f \sim 10^{-9}\,\mathrm{Hz}\,(M_{\rm PBH}/M_\odot)^{-1/2}$.
The $\mu$Hz regime maps onto PBH masses
$M_{\rm PBH}\sim[10^{-7},10^{-4}]\,M_\odot$, a range discussed in
connection with microlensing anomalies, but never targeted by a dedicated GW
experiment. Ref.~\cite{Blas:2026xws} has quantified this opportunity, showing that the resonant response of a highly eccentric Earth--satellite orbit can probe the low-mass end of
the window, see Fig.~\ref{fig:PBH}.

Two other features make this science case even more compelling. First, the frequencies at which the sensitivity peaks correspond to cosmic times close to the electroweak phase transition. If (part of) DM is made of these PBHs, \mis{}
will probe an epoch of cosmic history at scales interesting for particle physics. Second, the projected reach is directly relevant to existing observations: the candidate microlensing events reported
by OGLE~\cite{Niikura:2019kqi} and Subaru-HSC towards Andromeda~\cite{Sugiyama:2026kpv} are compatible with a PBH population of masses $M_{\rm PBH}\sim[10^{-8},10^{-4}]\,M_\odot$ at fractions with respect to the DM energy density $f_{\rm PBH}\sim10^{-2}$--$1$.
Ref.~\cite{Blas:2026xws} estimates \mis{} 
would suffice to either detect the associated SIGW background or exclude a primordial origin for these events. 

\begin{sciencebox}[Science goal SGB3 --- traces from primordial black holes]
\Mis{} will be sensitive to models primordial black holes of masses $M_{\rm PBH}\sim[10^{-7},10^{-4}]\,M_\odot$ where they constitute a substantial fraction of dark matter. 
\end{sciencebox}
\newpage

\subsection{Fundamental gravitation and fifth forces}\label{sec:fundamental}

Besides the main objectives of \mis{} from previous sections, its unique sensitivity to new forces that may act on its gravitational dynamics makes it a highly sensitive probe of possible extensions of general relativity and of the four known forces of Nature. Furthermore, some particles that may constitute the DM or whose existence may hint towards what completes the standard model of particle physics, or general relativity (GR), also generate new forces that may be sourced by Earth, see e.g.~\citep{moody:1984tq,fischbach:1999ly}.

\begin{wrapfigure}{R}{0.42\textwidth}
\vspace{-14pt}
  \centering
  \includegraphics[width=0.41\textwidth]{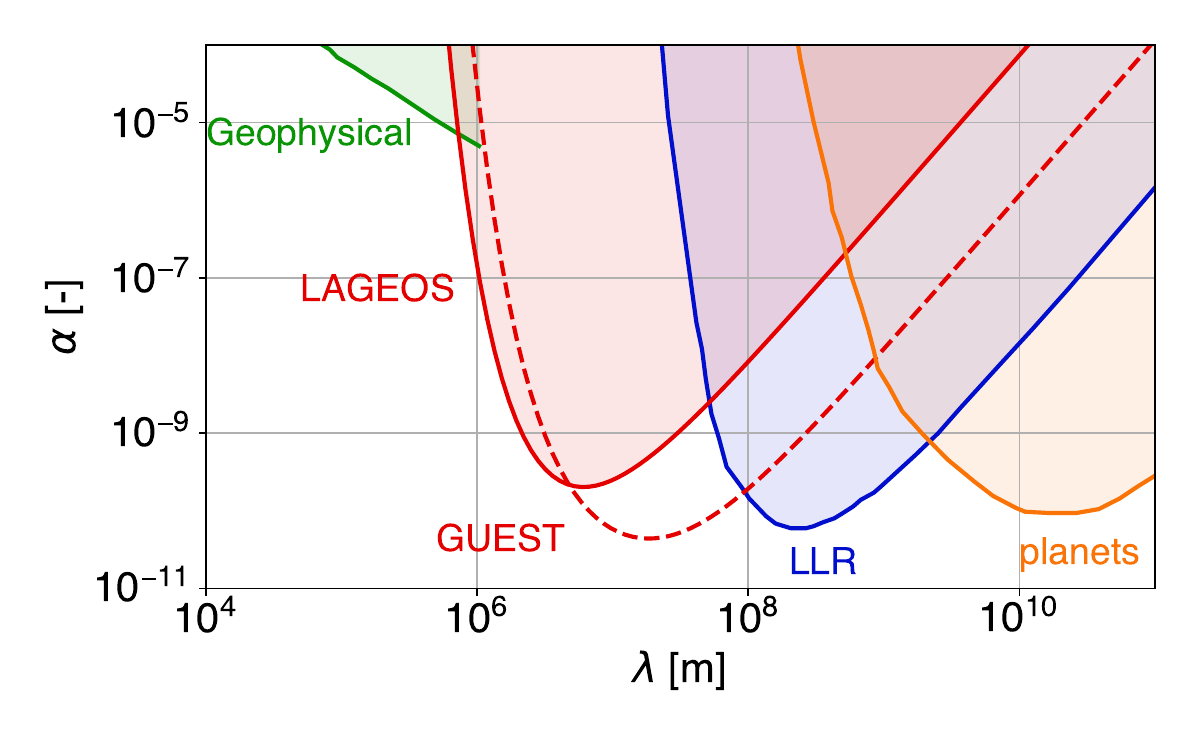}\vspace{-8pt}
  \caption{Existing constraints on a Yukawa fifth force (strength $\alpha$, range $\lambda$) and the region opened by \mis{}.}
  \label{fig:5thforce}
\vspace{-9pt}
\end{wrapfigure}

The novelty of tracking orbits with much higher eccentricity and longer periods than those of other satellites deployed for {\it precise} laser tracking (cf.~the \href{https://ilrs.gsfc.nasa.gov/missions/satellite_missions/current_missions/lars_general.html}{LARES}  and \href{https://lageos.gsfc.nasa.gov/}{LAGEOS}  satellites, which are in circular orbits from 1450 to 5800 km altitude) opens new possible searches for fundamental forces {\it sourced by the Earth}. In particular, it allows for a very precise measurement of the drift of the orbital argument of pericenter $\dot \omega$, which is highly sensitive to modifications to general relativity~\citep{2010PhRvL.105w1103L,Lucchesi:2014uza}, breaking of Lorentz symmetry~\citep{bailey:2006uq,lucchesi:2026aa} and new long-range forces ($5$th forces) which arise in models beyond the standard model of particle physics~\citep{moody:1984tq,fischbach:1999ly}. According to an analysis of LAGEOS data~\citep{Lucchesi2020}, the accuracy in $\dot \omega$ scales as $\sim1/(e P^{2/3})$. For \mis{} values, this translates into a $\sim30\times$ improvement in the determination of $\dot\omega$ with respect to LAGEOS, which can be used to measure relativistic effects such as the Schwarzschild precession~\citep{Lucchesi:2014uza} or constrain extensions of GR. The two-satellite configuration allows linear combinations of nodes and pericentres that cancel both $J_2$ and $J_4$ simultaneously, further isolating the relativistic signal from other effects.

A new particle with small mass and a coupling to ordinary matter generates a new long-range potential that, in the
simplest case, takes the Yukawa form
\begin{equation}
  V(r) = -\alpha\,\frac{G \Mearth}{r}\, e^{-r/\lambda},
  \qquad
  \lambda = \frac{\hbar}{m c}.
\end{equation}
The dimensionless strength $\alpha$ parameterises the new coupling; the range $\lambda$ is set by the mass of the mediator. Such forces arise generically in extensions of the
Standard Model: dilatons and moduli of string compactifications, light gauge bosons of
hidden $U(1)$s, Brans--Dicke-type scalar fields, and many ultra-light DM candidates
directly predict them~\citep{moody:1984tq,fischbach:1999ly}. Current constraints on $V$ and the possible reach of \mis{} are shown in Fig.~\ref{fig:5thforce}.  

\subsubsection{Quadratic coupling}

\begin{wrapfigure}{hR}{0.42\textwidth}
\vspace{-14pt}
  \centering
  \includegraphics[width=\linewidth]{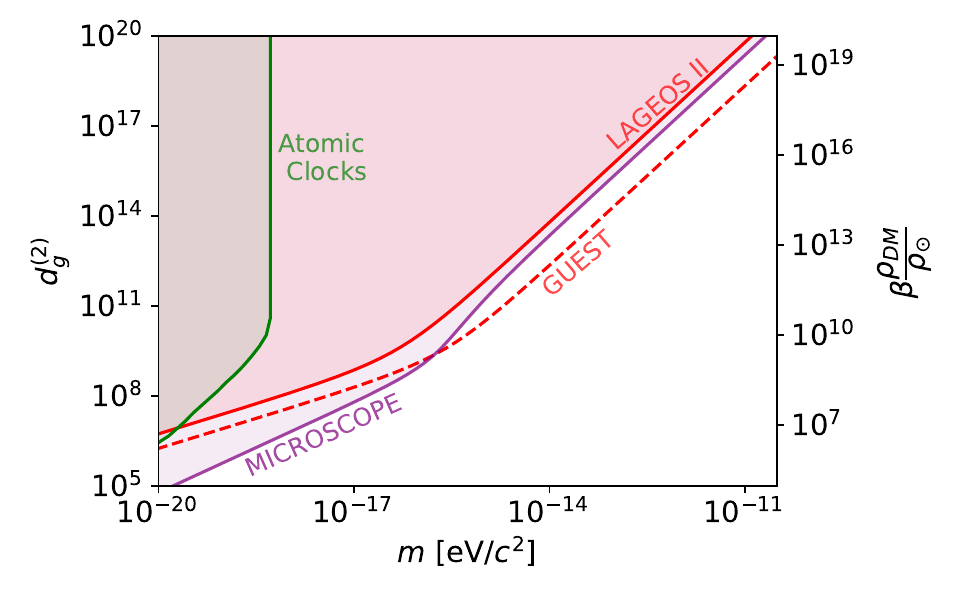}
  \vspace{-19pt}
    \caption{Existing constraints on the mass and coupling strength of ultra-light dark matter scalars with quadratic couplings to matter from existing experiments, and forecast constraints from measurements of  {\rm GUEST} perigee precession.}
    \label{fig:quadrcoup}
\vspace{-12pt}
\end{wrapfigure}

 A new scalar $\phi$ coupled quadratically to matter also induces scalar-mediated fifth forces on a test body with volume $V$ and density $\rho$ of the form:
\begin{equation}
    \vec{F}_{5}=-2\pi G \beta\int_V d^3x\,\rho\,\vec{\nabla}\phi^2\;,
\end{equation}
where the coupling parameter $\beta$ is dimensionless. These couplings are relevant for prominent models for dark matter, including axions \cite{Bauer:2023czj, Grossman:2025cov} and Higgs portal dark sectors \cite{OConnell:2006rsp,Patt:2006fw}. These fifth forces can be significantly suppressed in terrestrial searches \cite{Burrage:2025grx}, motivating a need for space-based experiments. Fig.~\ref{fig:quadrcoup} exemplifies forecast constraints on these models from measurements of the perigee precession rate of \mis{}, also determined by scaling the total error in the perigee precession rate as in the previous section. Current constraints on the parameter space of these models are also shown in Fig.~\ref{fig:quadrcoup} (see \cite{Burrage:2026loe, Hees:2018fpg}). The coupling parameter on the leftmost vertical axis is given in the notation of existing constraints, described in Ref.~\cite{Damour:2010rp}, with $d^{(2)}_g$ parametrising the coupling of the scalar to gluons. Corresponding values of a universal coupling  $ \beta\frac{\rho_{\rm DM}}{\rho_\odot}=\mathcal{O}(1)\times d_g^{(2)}$ for iron are given on the rightmost axis. Note that constraints shown from atomic clocks and MICROSCOPE necessarily depend on $\beta$ exhibiting composition dependence, and are absent if the coupling of the scalar to matter is universal.

\begin{sciencebox}[Science goal SGC --- fundamental physics]
\Mis{} will improve bounds on Yukawa fifth forces in an unexplored range of
$\lambda$, improve bounds on quadratic couplings, test relativistic precession at the few-percent level,  and provide competitive constraints on Lorentz-symmetry breaking in the gravitational sector.
\end{sciencebox}

\subsection{Geodesy, Earth science, and navigation}\label{sec:geodesy}

While \mis{} is designed first and foremost as a fundamental physics mission, it also offers compelling returns in the field of geodesy and its applications.

Geodesy---the measurement of Earth's orientation, shape, and gravity field---saw a revolution with the advent of the space era. The exquisite accuracy of space geodesy products makes them highly valuable to other fields in their own right, such as near-real-time observations of the state of the atmosphere, or detecting climate-change signals in the orientation of the Earth. Geodetic products such as the International Terrestrial and Celestial Reference Frames (ITRF and ICRF), and the Earth Orientation Parameters (EOP), are essential for virtually all space-based positioning and navigation systems. 

\vspace{-0.1cm}
SLR contributes to the development of Earth's gravity field models, the realisation of the ITRF, and supports science and navigation satellite missions. The two \href{https://lageos.gsfc.nasa.gov/}{LAGEOS}  ($a\simeq$12\,270\,km, $e\simeq0.0044$, and $a\simeq$12\,162\,km, $e\simeq0.0138$), and the two \href{https://ilrs.gsfc.nasa.gov/missions/satellite_missions/current_missions/lars_general.html}{LARES} satellites ($a\simeq$7\,820\,km, $e\simeq0.0012$, and $a\simeq$12\,266\,km, $e\simeq0.0003$), are the current workforce of the SLR space segment for geodetic purposes. Their contribution will be augmented by the ESA's mission \href{https://www.esa.int/Applications/Satellite_navigation/Mapping_planet_Earth_for_better_positioning_ESA_s_GENESIS_mission}{Genesis} (launch in 2028; see \citep{Delva2023}), to be tracked by SLR and other techniques.

\vspace{-0.1cm} 
New passive geodetic satellites in \textbf{\textit{highly eccentric ($e>0.5$)}} orbits and with \textbf{\textit{long periods ($P>$ day)}} provide an orbital regime currently unavailable to space geodesy. Most conveniently, additional geodetic products from \mis{} come from the same tracking data used for the primary science. 

\subsubsection{Dynamic scale: the standard gravitational parameter $GM_\oplus$.}

\begin{wrapfigure}{hR}{0.52\textwidth}
\vspace{-18pt}
\begin{center} 
    \includegraphics[width=\linewidth]{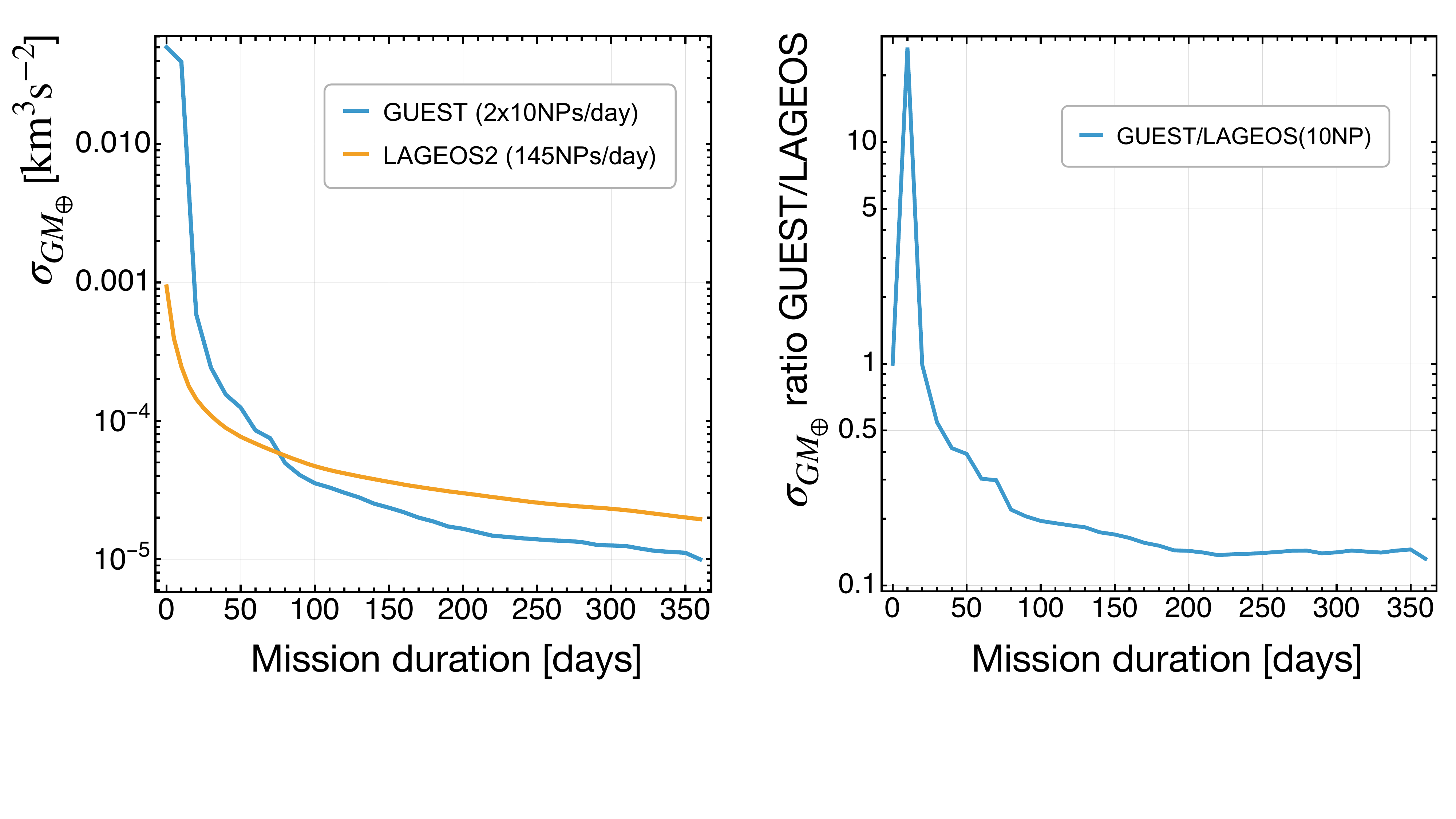}\hfill
  \end{center}\vspace{-25pt}
  \caption{Estimated formal errors in $GM_\oplus$ obtained from simulations of LAGEOS-2 and GUEST orbits. The simulations do not include additional parameters to be explored in future work. 
  \label{fig:gm_from_seb}
  }\vspace{-0.3cm}
\end{wrapfigure}

The first term in the multipole expansion of the Earth's potential is
$G\Mearth$, the standard gravitational parameter. This single number provides the
dynamical scale for every Earth-bound orbit and enters
every navigation solution, every satellite altimetry measurement, and every geodetic
frame. The currently adopted value, from~\cite{Ries:1992}, is known to a relative
precision of $\sim 2$ parts per billion.  The ambitious 1\,mm accuracy and 0.1\,mm/year stability goals of the Global Geodetic Observing System~\citep{Plag2009} will require a significant improvement in the performance of the observational networks, models, and processing strategies. In terms of accuracy, the current uncertainty in $GM_\oplus$ exceeds these goals by almost an order of magnitude.

Furthermore, despite repeated attempts to improve its error budget employing different combinations of geodetic satellites at different heights and new processing strategies, the currently available SLR space segment cannot provide further significant enhancements. The crux of the problem is that, for circular orbits, range biases, the up component of station position, and $GM_\oplus$, are all highly correlated, limiting the sensitivity of the orbits to each of these parameters. On the other hand, \mis{} orbits provide the means to achieve a step change improvement in the estimation of $GM_\oplus$, as their large eccentricity and semi-major axes allow for the effective decorrelation of the key parameters involved in the computation. From preliminary simulations, cf. Fig.~\ref{fig:gm_from_seb}, it is expected that after only a few months of tracking, \mis{} would match the uncertainty of the current $GM_\oplus$ standard, and after one year it would approach an uncertainty equivalent to 1\,mm on the surface of the Earth. This uncertainty reduction benefits all science applications that depend on an absolute knowledge of their position to accomplish their objectives (e.g.\,altimetry and geodesy), and it is a necessity for the Genesis mission, which aims to improve the state of the art in the realisation of the ITRF and the detection of inter-technique systematic errors at the level of 1\,mm.

\subsubsection{Augmenting the space segment and the ground network}

The realisation of the ITRF and computation of EOPs could benefit from having additional targets with sufficient tracking at orbital regimes previously unavailable. 
The space of orbital parameters sampled by these orbits may also benefit an improved estimation of the geopotential quadrupole, nowadays essentially computed from LAGEOS and LARES-2.
\mis{}, with a similar perigee, but sampling different inclinations, would be an additional contribution.

Finally, the approval of \mis{} may also generate a parallel effort in the extension and upgrade of the network of current laser ranging stations, at locations complementing current ranging capabilities. 
The availability of observations from a greater share of the network will increase their weight in the combined solutions aiming to generate geodetic products.

\begin{sciencebox}[Ancillary goal SGAnc --- geodesy and the GGOS]
\Mis{} adds an orbital regime unavailable to the present SLR space segment.
This enables a step change in the determination of $G\Mearth$, directly supporting
the Global Geodetic Observing System's 1-mm accuracy and 0.1-mm/year stability
goals, as well as the science objectives of GENESIS and future Earth-observation
missions.
\end{sciencebox}

\subsection{Synergies across communities}\label{sec:synergies}

\Mis{} sits at an intersection of scientific communities that do not often share an
experimental platform.  Besides fostering this multi-disciplinary dialogue, \mis{} brings distinct scientific opportunities for each of these communities:

\textbf{For the gravitational-wave community.}
\Mis{} uniquely fills the gap of four decades in frequency between PTAs and LISA, with a sensitivity to monochromatic strains of $h_0 \lesssim 10^{-19}$ and to stochastic
backgrounds $h_c \lesssim 10^{-17}$. The 2034--2044 operations window overlaps with
LISA, enabling multi-band follow-up of MBHB sources. A chirping source that starts
in the \mis{} band may cross into LISA within the mission lifetime, providing
\emph{certified} monochromatic calibration for LISA and possible improvements in sky localisation.

\textbf{For the particle-physics and primordial cosmology community.}
\Mis{} is a \emph{discovery machine} for new physics at energy scales beyond terrestrial
colliders. A first-order phase transition at $\sim$100 MeV, directly imprinted in the GW
spectrum in the $\muHz$ band, is precisely the regime that \mis{} covers. Ultra-light
bosons --- including dilatons, moduli, B$-$L gauge bosons, fuzzy dark matter, and string-theory
axions --- have coupling regions accessible \emph{only} through a mission such as
\mis{}. As a cosmological probe, it is sensitive to non-standard
inflation, first-order phase transitions, and cosmic-string networks. 
Fifth-force constraints at astronomical ranges are improved by a factor
$\sim 30$ over LAGEOS.

\textbf{For the astrophysics community.}
\Mis{} offers a diagnostic of the SMBHB population at $\muHz$ frequencies, bridging the
observation window between PTAs and LISA. It will probe the population of
intermediate-mass black holes in nearby globular clusters and, potentially, a solar-mass
companion of Sgr~A$^*$. Multi-messenger
follow-up of SMBHB candidates in the electromagnetic bands is an obvious opportunity that the mission is designed to enable.

\textbf{For the geodesy community.}
The
extension of the passive SLR space segment to high-eccentricity, long-period orbits
provides a new anchor for geopotential low-degree harmonics, and the ground-segment
upgrades benefit ITRF realisation, EOP determination, and virtually every space-based
navigation and Earth-observation programme.

\section{Mission implementation}\label{sec:implementation}

The scientific payoff of \mis{} rests on three pillars of implementation: the satellites,
the orbits, and the ground segment. For each of them, existing heritage is solid, which 
supports the idea of \mis{} as a mission to be implemented in the near future.

\subsection{The satellites: dense, passive spheres}

Each \mis{} satellite is a dense passive sphere covered in corner-cube retroreflectors
(CCRs). The considerations for their design are \textit{(i)} a large area is preferred, for more visibility; 
\textit{(ii)} small area vs mass ($A/M$) is preferred, to suppress the effects and uncertainties of non-gravitational accelerations\footnote{In particular, solar radiation pressure, Earth albedo and thermal re-radiation via the Yarkovsky--Schach effect.}
($A/M\sim 7 \times 10^{-4}\,\rm m^2/kg$ of LAGEOS is considered sufficient);
\textit{(iii)} low mass is preferred for a more efficient launch; \textit{(iv)} high conductivity is preferred, to suppress thermal noise; \textit{(v)} asymmetric design to guarantee gyroscopic stability. 

Two concepts are suggested, depending on other constraints to be considered in the final design: 

\begin{itemize}\setlength\itemsep{1pt}
\item \textbf{Concept A:} a monolithic sphere, with heritage from LARES-2 (2022). To allow for a large diameter, without excessive weight, Al~6061-T6 is suggested. As an example, a sphere of 86~cm in diameter and 845~kg is considered ($A/M\sim 7 \times 10^{-4}\,\rm m^2/kg$). 

\item \textbf{Concept B:} a hybrid design allows for additional flexibility to increase the area-to-mass ratio while keeping a low mass. 
Following the LAGEOS concept~\citep{visco2016}, a combination of an Al~6061-T6 outer (thick) shell, a tungsten-alloy core, and a Be--Cu central stud is suggested. A concept of 70~cm and 550~kg can be considered as a minimal possibility for visibility ($A/M\sim 7 \times 10^{-4}\,\rm m^2/kg$).
\end{itemize}

In both concepts, one can easily engineer an oblateness parameter $\gamma =(I_{zz} - I_{xx}) / I_{xx}
\approx 12.3\%$ (10.47\%) which guarantees sufficient gyroscopic stability.\footnote{For comparison, LAGEOS has an oblateness of $\sim$4\% and LARES of $\leq$0.2\%~\citep{visco2016}.} The higher value chosen for \mis{} ensures that the spin axis remains trapped along $z$ throughout the mission, simplifying the thermal and electromagnetic torque modelling. The spin evolution resulting from the interplay between gravitational precessional torques and electromagnetic braking will be modelled following the methodology of Visco \& Lucchesi~\citep{visco2018}, adapted for the \mis{} geometry and orbital environment.

\subsection{Cube Corner Retroreflectors}
\label{sec:ccrs}
 
\begin{figure}[H]
\centering
\begin{subfigure}[t]{0.42\textwidth}
  \includegraphics[width=0.92\linewidth]{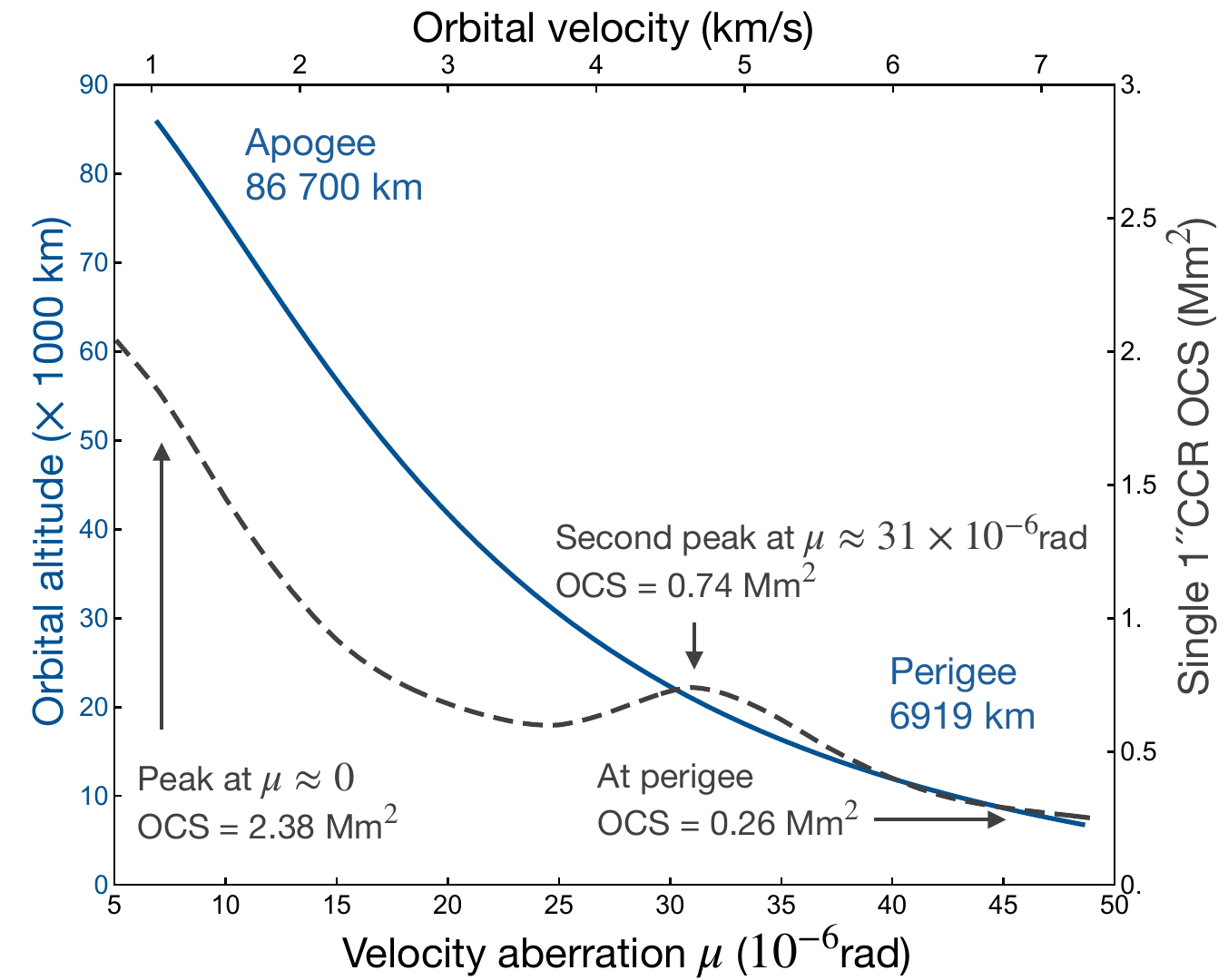}
  \caption{Optical cross-section coverage as a function of altitude.
  CCR1 and CCR2 populations ensure continuous SLR returns across the entire orbit.}
\end{subfigure}\hfill
\begin{subfigure}[t]{0.5\textwidth}
  \includegraphics[width=\linewidth]{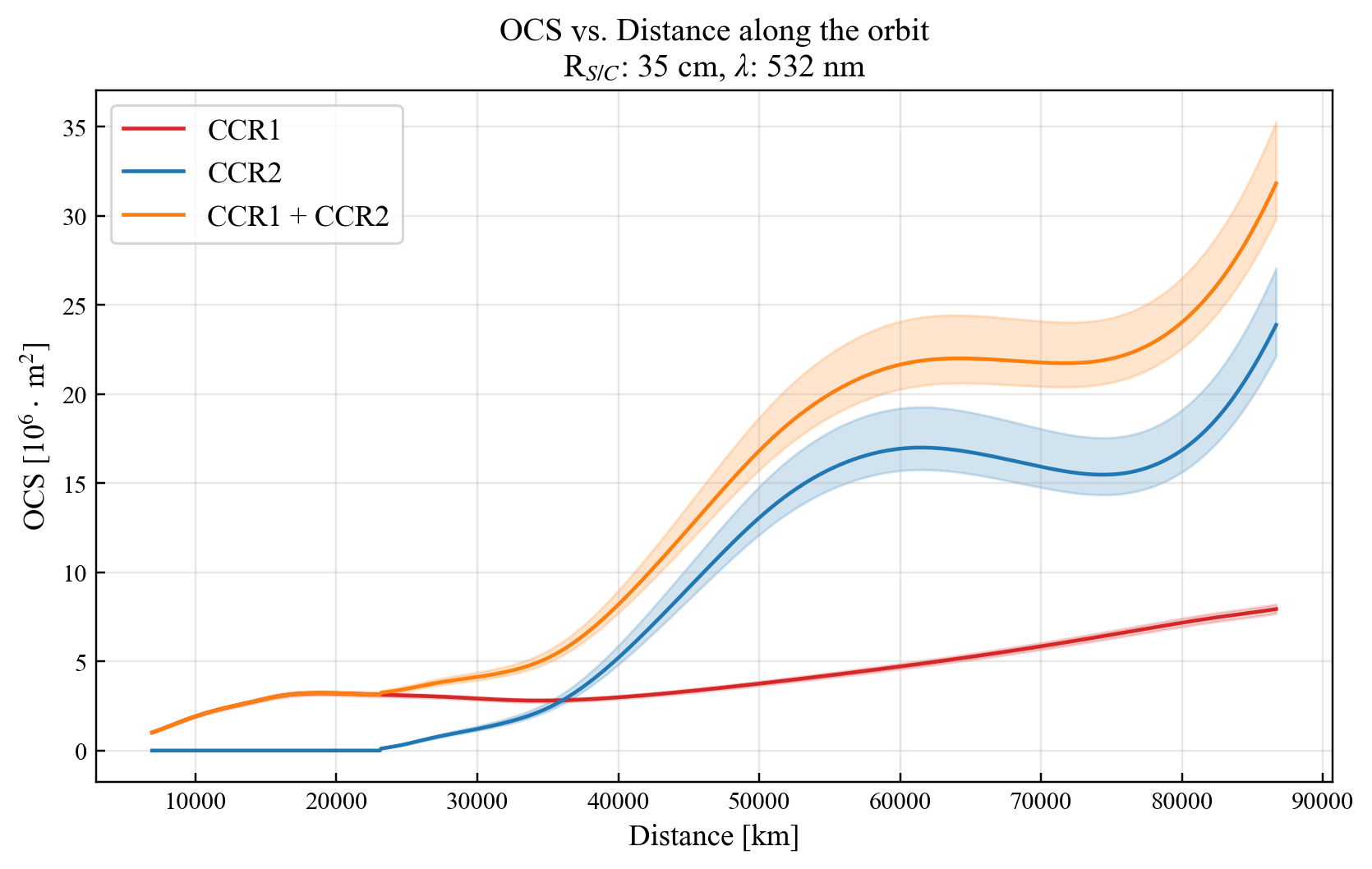}
  \caption{Visibility in $\lambda= 532$~nm  in terms of distance along the orbit. The red curve represents CCR1 average OCS, while the blue curve represents CCR2 average OCS. The total satellite OCS is given by the orange curve. The confidence bands represent the whole range of possible OCS according to the considered line of sight.}
\end{subfigure}
\caption{Estimated satellite’s OCS versus altitude for the \mis{} orbit (for the 70 cm sphere, factor 1.5 larger for 86 cm). }
\label{fig:OCS}
\end{figure}

 The retroreflector payload needs to generate sufficient optical cross-section (OCS, which is a measure of how visible an object is to a telescope) along the orbit. A new challenge, as compared to other geodesy missions, or even the Moon, is that the high eccentricity implies very different S/C velocities along the orbit: between 1.4~km/s (at apogee) and 9.5~km/s (at perigee). Such a difference in velocity translates into a difference in velocity aberration of several arcseconds between perigee and apogee; see the left panel from Fig.~\ref{fig:OCS}.
 
\mis{} addresses this issue through an innovative design based on two families of corner-cube reflectors mounted \emph{on the same sphere}.
\begin{itemize}
\item \textbf{CCR1} ({1} {inch}): optimised for the high velocity aberration $\mu \approx {48}\, {\mu \rm rad}$ encountered at perigee.  The far-field diffraction pattern (FFDP) of these prisms peaks as close as possible to this aberration value. This CCR also benefits from the LARES-2 \textit{heritage} of ASI.

\item \textbf{CCR2} ({2} {inch}): optimised for the low velocity aberration $\mu \approx {7}\, {\mu\rm rad}$ at apogee.  The larger aperture provides higher OCS per unit at small aberration, partially compensating the $R^4$ signal loss at extreme range.  
These CCR2s inherit from the 2-inch CCRs of COTS-type (Commercial Off-The-Shelf) acquired and tested by INFN for: (1) ESA’s \href{https://navisp.esa.int/project/details/211/show}{NAVISP} EL 1-61 R\&D on LCNS (Lunar Communication and Navigation Services); (2) an internal INFN R\&D for LCNS; (3) an R\&D for ASI.
\end{itemize}

\vspace{-0.1cm}
Both sets of CCR are uncoated fused-silica retroreflectors exploiting total internal reflection.  
The baseline for mounting the CCR on protrusions extending outward from the spherical surface to ensure that the reflective tips of the 2-inch CCRs and 1-inch CCRs are at the same radial distance. This is done to reduce time-detection biases and ambiguities in the calculation of the laser range correction and in the corresponding processing of the laser observations. Still, the advantage of this configuration with respect to no protrusions (like LAGEOS and LARES), or with CCR1s in recessions needs to be fully evaluated.

\begin{wrapfigure}{R}{0.4\textwidth}
\vspace{-30pt}
  \begin{center}
        \includegraphics[width=0.4\textwidth]{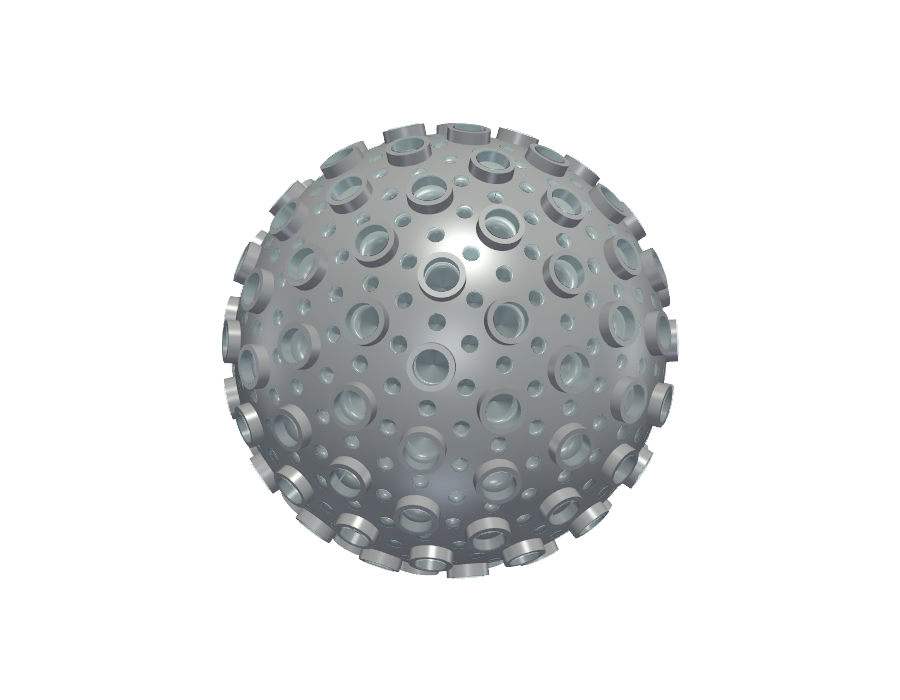}
  \end{center}
  \vspace{-30pt}
  \caption{Distribution of 1'' and 2 '' CCRs in the 3:1 configuration. Protrusions for CCR2 shown.
  \label{fig:SC3to1}
  }
  \vspace*{-.4cm}
\end{wrapfigure}
The distribution of the CCRs on the spheres has been optimized accounting for the following factors: \textit{(i)}  sufficient visibility; \textit{(ii)}    isotropy to provide constant visibility and minimize aspects such as thermal noise; \textit{(iii)}  small asymmetry to determine the spinning properties of the S/Cs; and \textit{(iv)}  sub-cm accuracy in the ranging measurements. The nominal solution is a (3:1) mix ratio (CCR1\,:\,CCR2); see Fig.~\ref{fig:SC3to1} and Table~\ref{tab:sat} for the exact numbers of CCRs, determined by a parametric optimisation of the total satellite OCS as a function of the number ratio. The weight of the CCR masses does not significantly alter the mass ranges we suggested for both concepts. 

The total OCS of the S/Cs are shown in Fig.~\ref{fig:OCS} and in the left panel of Fig.~\ref{fig:herlocs}, the right panel for the 70 cm sphere concept (for an 86 cm sphere, the ratio of OCS is approximately $(86/70)^2\approx 1.5$). In Fig.~\ref{fig:herlocs},  the lower part of the black band corresponds to the 70 cm sphere, while the upper one is for an 86 cm sphere, highlighting the small improvement in SLR tracking induced by concept A (larger spheres).

\begin{table}[H]
\centering
\begin{tabular}{|c|c|c|c|c|c|}
\hline
\rowcolor{gray!30}
Concept & $\emptyset$   & Weight (kg)&  \# CCR1/CCR2  &OCS range\\
\hline
A &86 cm  & 845
& 
561/189
& 
 $1.6$-$48.6$ Mm$^2$ \\
B&70 cm  & 525 
& 
366/124
& 
 $1.0$-$31.8$ Mm$^2$ \\
\bottomrule
\end{tabular}
\caption{Summary of main properties for the \mis{} satellites related to visibility.  The 4th column presents the estimated OCS of the two concepts.}
\label{tab:sat}
\end{table}

\subsection{Launch, deployment, and the orbital-transfer strategy}

A full exploration of the \mis{} launch strategy is beyond the scope of this work. Nonetheless, it is interesting to present
some of its peculiarities and which kinds of launchers could be appropriate. The $\Delta v$ needed to reach the two distinct
orbital planes for such heavy satellites (and for the platform that carries them to their final orbit, as was done for LARES-2) is significant. The nominal strategy is to perform a single launch, where the first satellite is directly injected by the upper stage of the launcher, while the second one is transferred to its orbit by an orbital-transfer vehicle (OTV). Two OTV options are possible: a
conventional chemical OTV using the BERTA hypergolic engine, and an electric variant based
on Hall-effect thrusters, offering higher specific impulse at the cost of longer
transfer time. Note that the large separation in the planes of the orbits of \mis{} (cf. Table~\ref{tab:orbits}) implies that the injection of the second orbit absorbs a lot of resources and may require non-trivial manoeuvres.

Another non-standard aspect of the launch is the need to impart a spin of 40--60~rpm to each
sphere after separation. The reason behind this is to ensure gyroscopic stability and an even thermal and
magnetic environment on orbit, which is key to achieving control over non-gravitational effects at the level required for the precise orbit determination we need in \mis~\cite{Lucchesi:2014uza}.

Within the \mis{} collaboration, we have identified several launch strategies within the boundaries of an F-class mission of ESA (carried by Ariane 6). Any other carrier with capacities similar to Ariane 6 would be equally adequate.

\subsection{The ground segment: the ILRS network, augmented}

The tracking of \mis{} will benefit from the extended expertise and coverage of the \href{https://ilrs.gsfc.nasa.gov/}{International Laser Ranging Service} (ILRS) network. Still, the peculiar orbits of the mission, of high eccentricity and large periods, present new challenges that we now describe. To put them into context, recall that our nominal target is 20 NPs observations per orbit for the whole network (10 per satellite), each with mm precision and cm accuracies, with integration times of about 2-10 minutes per NP. This is a modest number compared to the $O(100)$ points per day of LARES-2~\citep{Geisser2023,Sosnica2025a}  (see also the \href{https://edc.dgfi.tum.de/en/stations/}{EUROLAS Data Center} or the \href{https://ilrs.cddis.eosdis.nasa.gov/network/system_performance/global_report_cards/quarterly/index.html}{ILRS Quaterly analysis} or compare to \href{https://h2020nav.esa.int/project/h2020-038-10}{ESA-GASTON}).

\begin{figure}[H]
\centering
\begin{subfigure}[t]{0.45\linewidth}
\includegraphics[width=\linewidth]{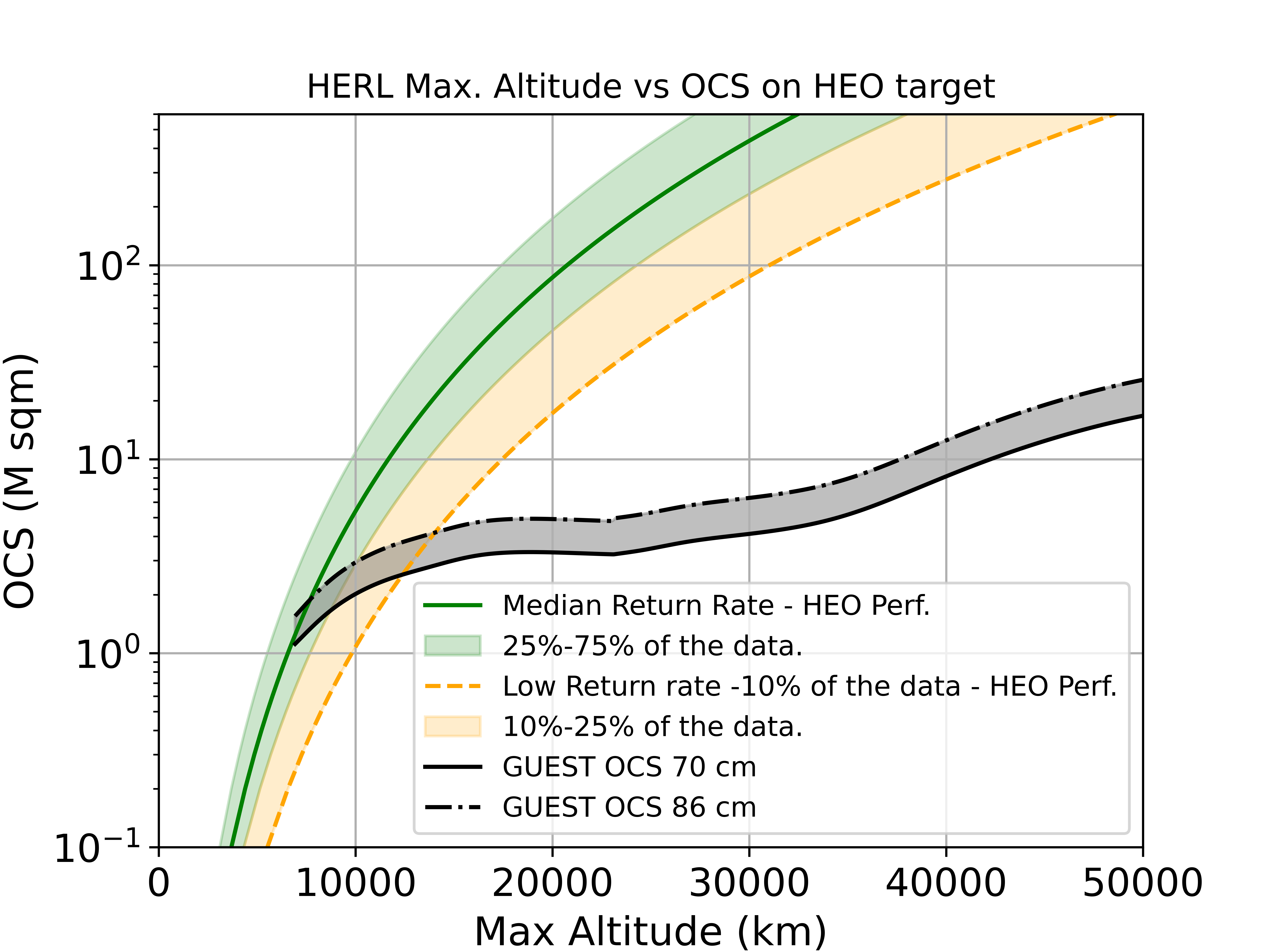}
\caption{Current high-end SLR station range performance.}
\end{subfigure}\hfill
\begin{subfigure}[t]{0.45\linewidth}
\includegraphics[width=\linewidth]{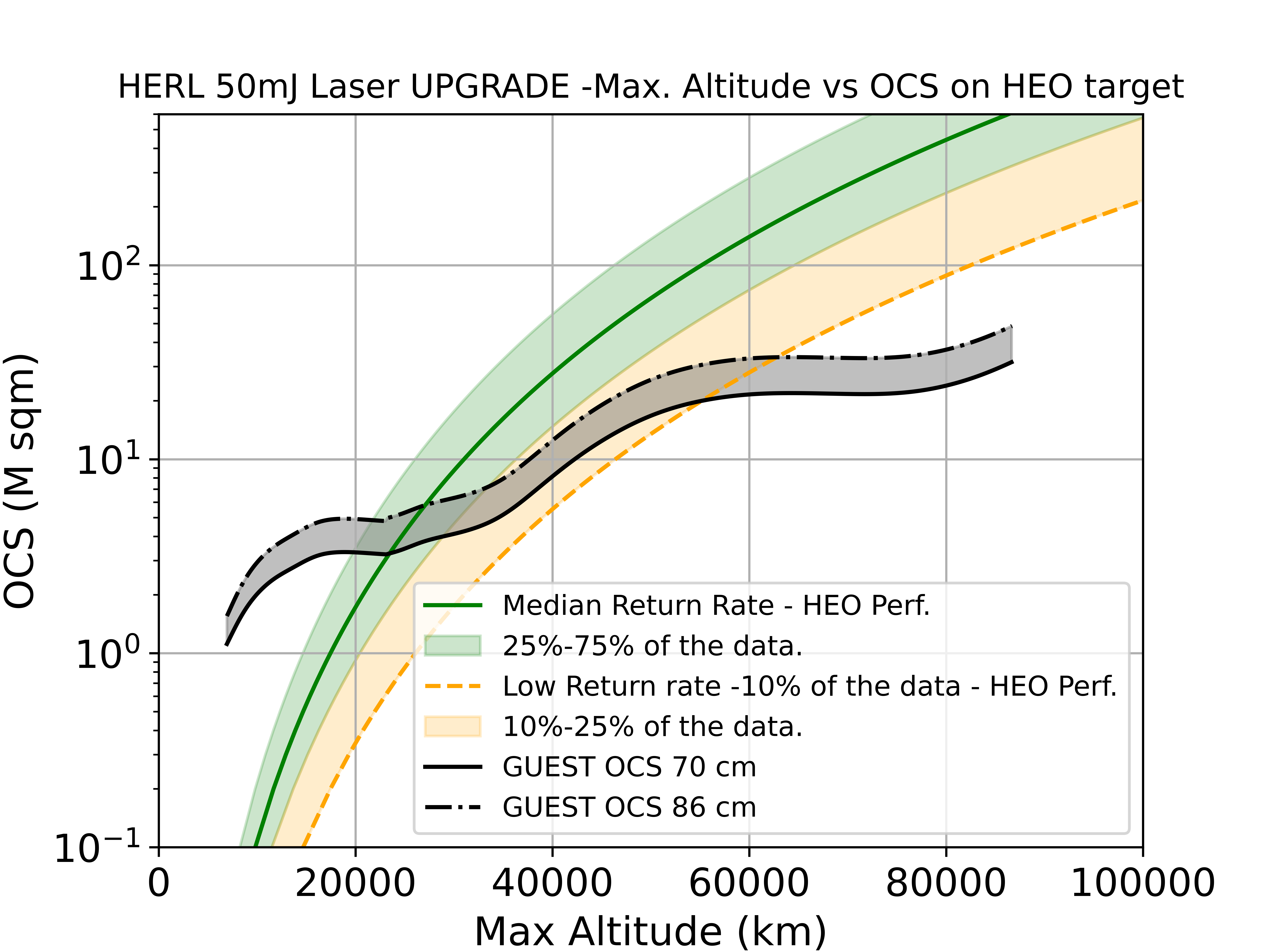}
\caption{Same station after the 50-mJ laser-power upgrade foreseen for \mis{}.}
\end{subfigure}
\caption{SLR station performance at current and upgraded laser power levels. The
upgrade provides the margin needed to track \mis{} at apogee while maintaining millimetre-level normal-point precision.}
\label{fig:herlocs}
\end{figure}
As compared to other geodesy satellites, the \mis{} satellites orbit at altitudes well above 10000~km most of the time, and travel at speeds that decrease their visibility when orbiting closer to the stations near perigee. To understand the implications of these facts, we will divide the laser ranging stations into those 
that are capable of ranging the Moon (LLR stations) and those that are not (SLR stations). The following LLR stations have been contacted by \mis{} collaboration, and have expressed their interest in contributing to the data acquisition: 
{\bf LLR}: Métrologie Optique - Grasse (France),  Italian Space Agency Matera (Italy), Geodätisches Observatorium  Wettzell (Germany); Kunming Yunnan Observatory (China). SLR stations are more numerous and are essentially guaranteed to contribute to ranging data acquisition if \mis{} receives a positive evaluation by the ILRS.

In Fig.~\ref{fig:herlocs}, left panel, we present our estimates of the capability of a high-performance SLR station to observe an object with a given visibility (OCS) on an eccentric orbit as a function of distance.\footnote{These estimates are based on the real data distribution from the Herstmonceux station.} The gray band shows the range of possible OCS values for \mis{} satellites as a function of distance. The result is clear: a standard SLR station can only observe the portion of the orbit near perigee, where the satellite spends very little time. A similar plot for an LLR station would show that it can track \mis{} satellites along the entire orbit. Since all LLR stations are located in the Northern hemisphere, we have designed the \mis{} orbits such that their apogees are also in the Northern hemisphere.  To understand the capacities of the whole ILRS network, we have computed the average time of contact for different stations during two characteristic years of the mission, defined as the time over which each of the satellites is visible to a station in the network. The geographic distribution of these stations works in our favour: even when the perigee moves out of the visibility range of one station, their global coverage ensures that at least one station remains capable of tracking the orbit almost continuously (see Fig.~\ref{fig:req_tr}). We conclude that the 10 NPs per orbit that are required by \mis{} are \textit{realistically achievable by combining tracking by LLR stations above 14000 km with coverage by SLR stations near perigee}. We have also verified that uneven data cadence (having data clustered at different regions of the orbit) does not significantly affect the extraction of the signals targeted by \mis{}.

\subsubsection{Station upgrades and new stations}

Despite having argued that the current capacity of the ILRS is enough to provide the data required for \mis{} to achieve its scientific objectives, constraints from operations, weather, service, etc., may affect this tracking. This motivates the consideration of mitigation measures: upgrading and extension of the network, as we describe below.

\begin{figure}[H]
    \centering
        \vspace{-4pt}
        \includegraphics[width=\textwidth]{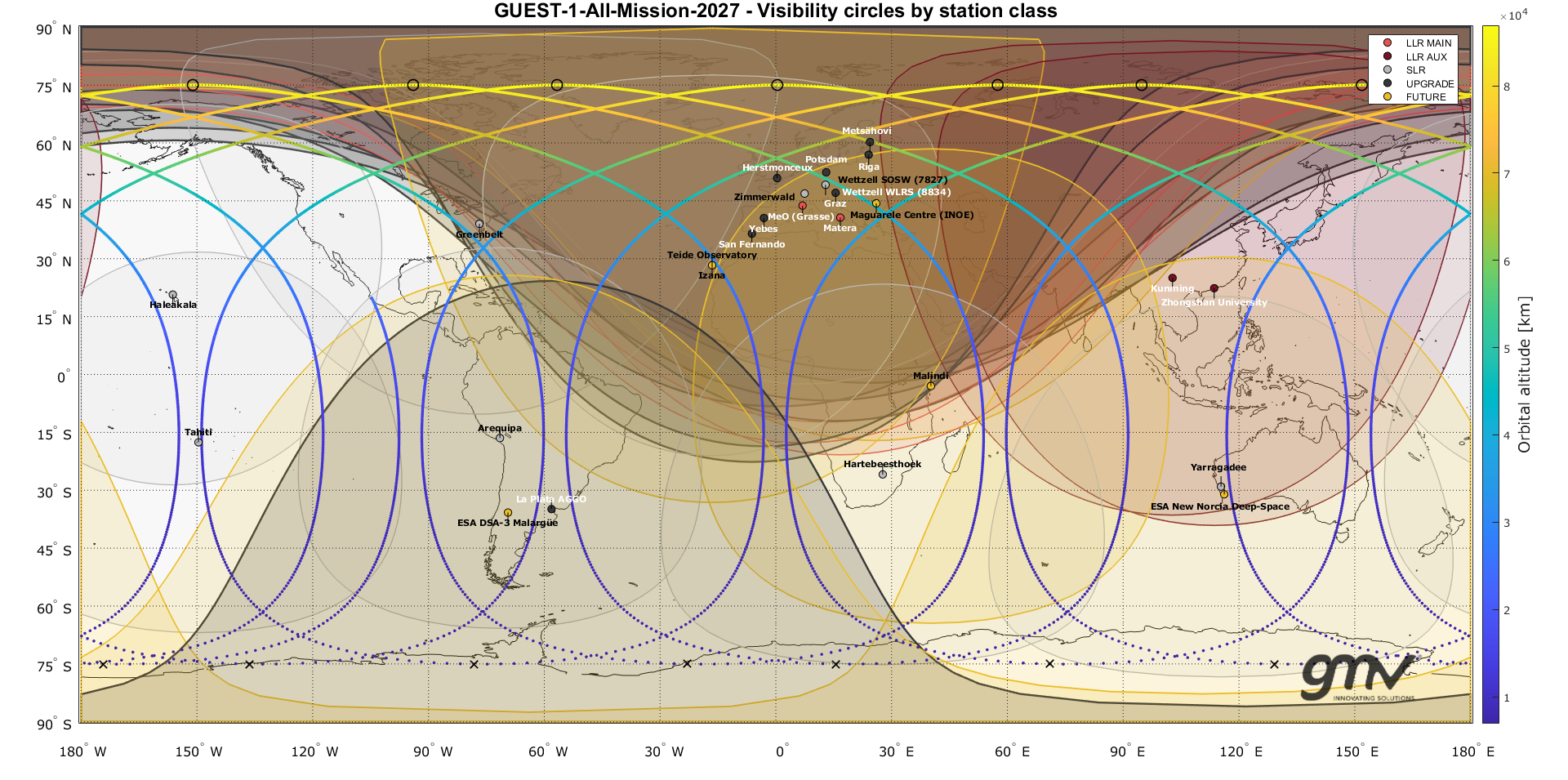}\vspace{-.3cm}
    \caption{Geometric visibility of the GUEST-1 orbits from different ranging stations. The continuous colored line represents the GUEST-1 orbit (the color corresponding to its altitude). Note that the orbit ground tracks will change over the mission's lifetime. The dashed areas represent the visibility areas for the different stations shown. They are classified as LLR (main stations are those with relations to people in this proposal), SLR, reach of possible upgrades to existing stations, and four possible new ones (appearing as FUTURE).
    \label{fig:req_tr}}\vspace{-0.3cm}
\end{figure}

\begin{itemize}\setlength\itemsep{1pt}
\item \textbf{Station upgrades.} Upgrading selected SLR
stations to a $\sim$50~mJ laser or telescopes on the order of 80 cm to 1 m diameter could extend their ranging reach up to 60000 km, as shown in the right panel of 
Fig.~\ref{fig:herlocs}. The cost of such improvements is modest and will also benefit other station activities (such as debris tracking). 

\item \textbf{New stations.} The fact that stations with demonstrated LLR capacities only exist in the Northern Hemisphere and at similar latitudes motivates the consideration of the deployment of new stations. This may be relevant not only to track \mis{} satellites, but also to support other scientific missions and more applied endeavors.
The ideal locations for \mis{} are those complementary to those of current LLR stations: Australia, the Pacific, South Africa, or South America (see Fig.~\ref{fig:req_tr}). 
\end{itemize}

\subsection{Mission timeline}

One of the objectives of \mis{} is to present a mission that can be launched within 3-5 years of approval. Once launched, the first spacecraft is almost immediately available, while the second one will require navigation to its planned orbit over a timescale of days to weeks. From that moment, both spacecraft will be fully passive, and the mission will rely on laser ranging data from Earth. The first milestones for the scientific objectives will be achieved in 10 years, with continued improvements possible past this point via the ongoing collection of NP data. After $\sim$30 years, both satellites will reenter.

\section{Summary and outlook}\label{sec:summary}

\Mis{} is a mission concept that delivers transformative science with an unusually simple architecture: two passive, retroreflector-covered spheres, on high-eccentricity Earth orbits, tracked by laser ranging from the ground over three decades. Such a concept fits within the financial boundaries of small missions, such as ESA F-class or NASA SMEX. 

The scientific case rests on a single physical insight: \emph{that a binary system may behave as a resonant detector of perturbations of fundamental origin}. From this follows a broad scientific program:
\begin{itemize}[itemsep=1pt]
\item first phenomenologically-relevant access to the gravitational-wave spectrum in the microhertz band,
filling a persistent gap between pulsar timing arrays and LISA;
\item first-ever sensitivity to individual supermassive-black-hole binary sources in this band, together with intermediate-mass BH candidates in nearby globular clusters;
\item a unique probe of early-Universe phenomena at energy scales inaccessible to
terrestrial colliders, including first-order phase transitions and cosmic-string
networks;
\item direct detection of ultra-light dark matter in an unexplored region of parameter
space;
\item gravitational-wave signatures of superradiant boson clouds around nearby
supermassive black holes, in a mass range accessible to no other detector;
\item possible detection of the consequences of planetary-mass primordial black holes;
\item improvement in fifth-force bounds at astronomical ranges;
and
\item a step change in the absolute determination of $G\Mearth$, with direct impact on the Global Geodetic Observing System.
\end{itemize}

The technical concept draws on four decades of SLR and LLR heritage and is mature enough to conceive a mission starting operations in 
4-5 years. The launch vehicle, the separation system, the
spin unit, and the optical design may be based on flight-proven subsystems. The ground
segment builds on existing ILRS stations with targeted upgrades and possible new stations that would benefit the entire
ILRS community.

A launch in the window 2034--2037 is not merely feasible but strategically compelling.
The overlap with LISA opens multi-band GW astronomy; the
overlap with GENESIS delivers the geodesy case; and the overlap with pulsar-timing arrays
provides the frequency-space lever arm needed to disentangle the recent evidence for
nanohertz GWs.

We hope that the broad community of particle physicists, gravitational-wave astronomers, astrophysicists, cosmologists, and geodesists shares our excitement about the milestones that \mis{} may achieve, and we invite all our colleagues to join the quest of building the necessary expertise, from science studies to hardware and data analysis, required to unveil the mysteries of the microhertz gravitational-wave sky.

\section*{Acknowledgements}
\addcontentsline{toc}{section}{Acknowledgements}

Logo designed by \`Eve Barlier. 

IFAE is partially funded by the CERCA program of the Generalitat de Catalunya.
This publication is part of the R\&D\&i project PID2023-146686NB-C31 funded by MICIU/AEI/10.13039/501100011033/ and by ERDF/EU.
D.Blas acknowledges financial support from the Spanish Ministry of Science and Innovation (MICINN) through the Spanish State Research Agency, under Severo Ochoa Centres of Excellence Programme 2025-2029 (CEX2024001442-S).
This work is supported by ERC grant (GravNet, ERC-2024-SyG 101167211, DOI: 10.3030/101167211). Funded by the European Union. Views and opinions expressed are, however those of the author(s) only and do not necessarily reflect those of the European Union or the European Research Council Executive Agency. Neither the European Union nor the granting authority can be held responsible for them. 
D.Blas acknowledges the support from the European Research Area (ERA) via the UNDARK project of the Widening participation and spreading excellence programme (project number 101159929).
INFN authors acknowledge the support from the Italian Space Agency (ASI) through the research Agreement ASI-INFN n. 2019-15-HH.0, through the research Agreement ASI n. 2022‐8‐HH.0 (science studies for ESA's Hera mission) and the support from the European Space Agency (ESA) through the ESA-INFN Contract n. 4000133721/21/NL/CR.
A.~C.~Jenkins was supported by the UK Engineering and Physical Sciences Research Council through a Stephen Hawking Fellowship (Grant No. EP/U536684/1) and by a Gavin Boyle Fellowship at the Kavli Institute for Cosmology, Cambridge.
A.~Hees, A.~Perego and N.~Tamanini acknowledge support from the French space agency CNES.
C.~Burrage is supported by an STFC Consolidated Grant  [Grant Nos.  ST/X000672/1]. A.~Macdonald
is supported by  EPSRC Quantum Technologies Doctoral Training Partnership (DTP) 2024-25. 
C.F.~Sopuerta is supported by contract PID2022-137674NB-I00/AEI/10.13039/501100011033 (Spanish Ministry of Science and Innovation). This work was also partly supported by the Spanish program Unidad de Excelencia María de Maeztu CEX2020-001058-M, financed by MCIN/AEI/10.13039/501100011033, and by the MaX-CSIC Excellence Award MaX4-SOMMA-ICE.
X.Xue is funded by the grant CNS2023-143767. Grant CNS2023-143767 is funded by MICIU/AEI/10.13039/501100011033 and by European Union NextGenerationEU/PRTR.
This publication is part of the grant CEX2024-001441-S, financed by MICIU/AEI/10.13039/501100011033/.


\phantomsection
\addcontentsline{toc}{section}{References}
\bibliography{refs}

\begin{thebibliography}{56}%
\makeatletter
\providecommand \@ifxundefined [1]{%
 \@ifx{#1\undefined}
}%
\providecommand \@ifnum [1]{%
 \ifnum #1\expandafter \@firstoftwo
 \else \expandafter \@secondoftwo
 \fi
}%
\providecommand \@ifx [1]{%
 \ifx #1\expandafter \@firstoftwo
 \else \expandafter \@secondoftwo
 \fi
}%
\providecommand \natexlab [1]{#1}%
\providecommand \enquote  [1]{``#1''}%
\providecommand \bibnamefont  [1]{#1}%
\providecommand \bibfnamefont [1]{#1}%
\providecommand \citenamefont [1]{#1}%
\providecommand \href@noop [0]{\@secondoftwo}%
\providecommand \href [0]{\begingroup \@sanitize@url \@href}%
\providecommand \@href[1]{\@@startlink{#1}\@@href}%
\providecommand \@@href[1]{\endgroup#1\@@endlink}%
\providecommand \@sanitize@url [0]{\catcode `\\12\catcode `\$12\catcode
  `\&12\catcode `\#12\catcode `\^12\catcode `\_12\catcode `\%12\relax}%
\providecommand \@@startlink[1]{}%
\providecommand \@@endlink[0]{}%
\providecommand \url  [0]{\begingroup\@sanitize@url \@url }%
\providecommand \@url [1]{\endgroup\@href {#1}{\urlprefix }}%
\providecommand \urlprefix  [0]{URL }%
\providecommand \Eprint [0]{\href }%
\providecommand \doibase [0]{https://doi.org/}%
\providecommand \selectlanguage [0]{\@gobble}%
\providecommand \bibinfo  [0]{\@secondoftwo}%
\providecommand \bibfield  [0]{\@secondoftwo}%
\providecommand \translation [1]{[#1]}%
\providecommand \BibitemOpen [0]{}%
\providecommand \bibitemStop [0]{}%
\providecommand \bibitemNoStop [0]{.\EOS\space}%
\providecommand \EOS [0]{\spacefactor3000\relax}%
\providecommand \BibitemShut  [1]{\csname bibitem#1\endcsname}%
\let\auto@bib@innerbib\@empty
\bibitem [{\citenamefont {Abbott}\ \emph {et~al.}(2016)\citenamefont {Abbott}
  \emph {et~al.}}]{LIGOScientific:2016aoc}%
  \BibitemOpen
  \bibfield  {author} {\bibinfo {author} {\bibfnamefont {B.~P.}\ \bibnamefont
  {Abbott}} \emph {et~al.} (\bibinfo {collaboration} {LIGO Scientific,
  Virgo}),\ }\href {https://doi.org/10.1103/PhysRevLett.116.061102} {\bibfield
  {journal} {\bibinfo  {journal} {Phys. Rev. Lett.}\ }\textbf {\bibinfo
  {volume} {116}},\ \bibinfo {pages} {061102} (\bibinfo {year} {2016})},\
  \Eprint {https://arxiv.org/abs/1602.03837} {arXiv:1602.03837 [gr-qc]}
  \BibitemShut {NoStop}%
\bibitem [{\citenamefont {Antoniadis}\ \emph {et~al.}(2023)\citenamefont
  {Antoniadis} \emph {et~al.}}]{EPTA:2023fyk}%
  \BibitemOpen
  \bibfield  {author} {\bibinfo {author} {\bibfnamefont {J.}~\bibnamefont
  {Antoniadis}} \emph {et~al.} (\bibinfo {collaboration} {EPTA, InPTA}),\
  }\href {https://doi.org/10.1051/0004-6361/202346844} {\bibfield  {journal}
  {\bibinfo  {journal} {Astron. Astrophys.}\ }\textbf {\bibinfo {volume}
  {678}},\ \bibinfo {pages} {A50} (\bibinfo {year} {2023})},\ \Eprint
  {https://arxiv.org/abs/2306.16214} {arXiv:2306.16214 [astro-ph.HE]}
  \BibitemShut {NoStop}%
\bibitem [{\citenamefont {Reardon}\ \emph {et~al.}(2023)\citenamefont {Reardon}
  \emph {et~al.}}]{Reardon:2023gzh}%
  \BibitemOpen
  \bibfield  {author} {\bibinfo {author} {\bibfnamefont {D.~J.}\ \bibnamefont
  {Reardon}} \emph {et~al.},\ }\href {https://doi.org/10.3847/2041-8213/acdd02}
  {\bibfield  {journal} {\bibinfo  {journal} {Astrophys. J. Lett.}\ }\textbf
  {\bibinfo {volume} {951}},\ \bibinfo {pages} {L6} (\bibinfo {year} {2023})},\
  \Eprint {https://arxiv.org/abs/2306.16215} {arXiv:2306.16215 [astro-ph.HE]}
  \BibitemShut {NoStop}%
\bibitem [{\citenamefont {Agazie}\ \emph {et~al.}(2023)\citenamefont {Agazie}
  \emph {et~al.}}]{NANOGrav:2023gor}%
  \BibitemOpen
  \bibfield  {author} {\bibinfo {author} {\bibfnamefont {G.}~\bibnamefont
  {Agazie}} \emph {et~al.} (\bibinfo {collaboration} {NANOGrav}),\ }\href
  {https://doi.org/10.3847/2041-8213/acdac6} {\bibfield  {journal} {\bibinfo
  {journal} {Astrophys. J. Lett.}\ }\textbf {\bibinfo {volume} {951}},\
  \bibinfo {pages} {L8} (\bibinfo {year} {2023})},\ \Eprint
  {https://arxiv.org/abs/2306.16213} {arXiv:2306.16213 [astro-ph.HE]}
  \BibitemShut {NoStop}%
\bibitem [{\citenamefont {Xu}\ \emph {et~al.}(2023)\citenamefont {Xu} \emph
  {et~al.}}]{Xu:2023wog}%
  \BibitemOpen
  \bibfield  {author} {\bibinfo {author} {\bibfnamefont {H.}~\bibnamefont {Xu}}
  \emph {et~al.},\ }\href {https://doi.org/10.1088/1674-4527/acdfa5} {\bibfield
   {journal} {\bibinfo  {journal} {Res. Astron. Astrophys.}\ }\textbf {\bibinfo
  {volume} {23}},\ \bibinfo {pages} {075024} (\bibinfo {year} {2023})},\
  \Eprint {https://arxiv.org/abs/2306.16216} {arXiv:2306.16216 [astro-ph.HE]}
  \BibitemShut {NoStop}%
\bibitem [{ESAVoyage2050()}]{Voyage2050}%
  \BibitemOpen
  ESAVoyage2050,\ \href@noop {} {\bibinfo {title} {{ESA Voyage 2050}}},\
  \bibinfo {howpublished}
  {\url{https://www.cosmos.esa.int/web/voyage-2050}}\BibitemShut {NoStop}%
\bibitem [{\citenamefont {{LISA collaboration}}(2024)}]{LISA:2024hlh}%
  \BibitemOpen
  \bibfield  {author} {\bibinfo {author} {\bibnamefont {{LISA
  collaboration}}},\ }\href@noop {} {\bibfield  {journal} {\bibinfo  {journal}
  {arXiv}\ } (\bibinfo {year} {2024})},\ \Eprint
  {https://arxiv.org/abs/2402.07571} {arXiv:2402.07571 [astro-ph.CO]}
  \BibitemShut {NoStop}%
\bibitem [{\citenamefont {Abdalla}\ \emph {et~al.}(2025)\citenamefont {Abdalla}
  \emph {et~al.}}]{Abdalla:2024sst}%
  \BibitemOpen
  \bibfield  {author} {\bibinfo {author} {\bibfnamefont {A.}~\bibnamefont
  {Abdalla}} \emph {et~al.},\ }\href
  {https://doi.org/10.1140/epjqt/s40507-025-00344-3} {\bibfield  {journal}
  {\bibinfo  {journal} {EPJ Quant. Technol.}\ }\textbf {\bibinfo {volume}
  {12}},\ \bibinfo {pages} {42} (\bibinfo {year} {2025})},\ \Eprint
  {https://arxiv.org/abs/2412.14960} {arXiv:2412.14960 [hep-ex]} \BibitemShut
  {NoStop}%
\bibitem [{\citenamefont {Kawamura}\ \emph {et~al.}(2011)\citenamefont
  {Kawamura} \emph {et~al.}}]{Kawamura:2011zz}%
  \BibitemOpen
  \bibfield  {author} {\bibinfo {author} {\bibfnamefont {S.}~\bibnamefont
  {Kawamura}} \emph {et~al.},\ }\href
  {https://doi.org/10.1088/0264-9381/28/9/094011} {\bibfield  {journal}
  {\bibinfo  {journal} {Class. Quant. Grav.}\ }\textbf {\bibinfo {volume}
  {28}},\ \bibinfo {pages} {094011} (\bibinfo {year} {2011})}\BibitemShut
  {NoStop}%
\bibitem [{\citenamefont {Amaral}\ \emph {et~al.}(2026)\citenamefont {Amaral}
  \emph {et~al.}}]{Amaral:2026bef}%
  \BibitemOpen
  \bibfield  {author} {\bibinfo {author} {\bibfnamefont {D.}~\bibnamefont
  {Amaral}} \emph {et~al.},\ }\Eprint {https://arxiv.org/abs/2603.24645}
  {arXiv:2603.24645 [astro-ph.IM]}  (\bibinfo {year} {2026})\BibitemShut
  {NoStop}%
\bibitem [{\citenamefont {Aggarwal}\ \emph {et~al.}(2025)\citenamefont
  {Aggarwal} \emph {et~al.}}]{Aggarwal:2025noe}%
  \BibitemOpen
  \bibfield  {author} {\bibinfo {author} {\bibfnamefont {N.}~\bibnamefont
  {Aggarwal}} \emph {et~al.},\ }\href
  {https://doi.org/10.1007/s41114-025-00060-5} {\bibfield  {journal} {\bibinfo
  {journal} {Living Rev. Rel.}\ }\textbf {\bibinfo {volume} {28}},\ \bibinfo
  {pages} {10} (\bibinfo {year} {2025})},\ \Eprint
  {https://arxiv.org/abs/2501.11723} {arXiv:2501.11723 [gr-qc]} \BibitemShut
  {NoStop}%
\bibitem [{\citenamefont {Sesana}\ \emph {et~al.}(2021)\citenamefont {Sesana}
  \emph {et~al.}}]{Sesana:2019vho}%
  \BibitemOpen
  \bibfield  {author} {\bibinfo {author} {\bibfnamefont {A.}~\bibnamefont
  {Sesana}} \emph {et~al.},\ }\href
  {https://doi.org/10.1007/s10686-021-09709-9} {\bibfield  {journal} {\bibinfo
  {journal} {Exper. Astron.}\ }\textbf {\bibinfo {volume} {51}},\ \bibinfo
  {pages} {1333} (\bibinfo {year} {2021})}\BibitemShut {NoStop}%
\bibitem [{\citenamefont {{Vaglio}}\ \emph {et~al.}(2025)\citenamefont
  {{Vaglio}}, \citenamefont {{Falxa}}, \citenamefont {{Mentasti}},
  \citenamefont {{Renzini}}, \citenamefont {{Kuntz}}, \citenamefont
  {{Barausse}}, \citenamefont {{Contaldi}},\ and\ \citenamefont
  {{Sesana}}}]{2025arXiv250718593V}%
  \BibitemOpen
  \bibfield  {author} {\bibinfo {author} {\bibfnamefont {M.}~\bibnamefont
  {{Vaglio}}}, \bibinfo {author} {\bibfnamefont {M.}~\bibnamefont {{Falxa}}},
  \bibinfo {author} {\bibfnamefont {G.}~\bibnamefont {{Mentasti}}}, \bibinfo
  {author} {\bibfnamefont {A.~I.}\ \bibnamefont {{Renzini}}}, \bibinfo {author}
  {\bibfnamefont {A.}~\bibnamefont {{Kuntz}}}, \bibinfo {author} {\bibfnamefont
  {E.}~\bibnamefont {{Barausse}}}, \bibinfo {author} {\bibfnamefont
  {C.}~\bibnamefont {{Contaldi}}},\ and\ \bibinfo {author} {\bibfnamefont
  {A.}~\bibnamefont {{Sesana}}},\ }\href
  {https://doi.org/10.48550/arXiv.2507.18593} {\bibfield  {journal} {\bibinfo
  {journal} {arXiv e-prints}\ ,\ \bibinfo {eid} {arXiv:2507.18593}} (\bibinfo
  {year} {2025})},\ \Eprint {https://arxiv.org/abs/2507.18593}
  {arXiv:2507.18593 [gr-qc]} \BibitemShut {NoStop}%
\bibitem [{\citenamefont {Hui}\ \emph {et~al.}(2013)\citenamefont {Hui} \emph
  {et~al.}}]{Hui:2012yp}%
  \BibitemOpen
  \bibfield  {author} {\bibinfo {author} {\bibfnamefont {L.}~\bibnamefont
  {Hui}} \emph {et~al.},\ }\href {https://doi.org/10.1103/PhysRevD.87.084009}
  {\bibfield  {journal} {\bibinfo  {journal} {Phys. Rev. D}\ }\textbf {\bibinfo
  {volume} {87}},\ \bibinfo {pages} {084009} (\bibinfo {year}
  {2013})}\BibitemShut {NoStop}%
\bibitem [{\citenamefont {Blas}\ and\ \citenamefont
  {Jenkins}(2022{\natexlab{a}})}]{Blas:2021mpc}%
  \BibitemOpen
  \bibfield  {author} {\bibinfo {author} {\bibfnamefont {D.}~\bibnamefont
  {Blas}}\ and\ \bibinfo {author} {\bibfnamefont {A.~C.}\ \bibnamefont
  {Jenkins}},\ }\href {https://doi.org/10.1103/PhysRevD.105.064021} {\bibfield
  {journal} {\bibinfo  {journal} {Phys. Rev. D}\ }\textbf {\bibinfo {volume}
  {105}},\ \bibinfo {pages} {064021} (\bibinfo {year}
  {2022}{\natexlab{a}})}\BibitemShut {NoStop}%
\bibitem [{\citenamefont {Blas}\ and\ \citenamefont
  {Jenkins}(2022{\natexlab{b}})}]{Blas:2021mqw}%
  \BibitemOpen
  \bibfield  {author} {\bibinfo {author} {\bibfnamefont {D.}~\bibnamefont
  {Blas}}\ and\ \bibinfo {author} {\bibfnamefont {A.~C.}\ \bibnamefont
  {Jenkins}},\ }\href {https://doi.org/10.1103/PhysRevLett.128.101103}
  {\bibfield  {journal} {\bibinfo  {journal} {Phys. Rev. Lett.}\ }\textbf
  {\bibinfo {volume} {128}},\ \bibinfo {pages} {101103} (\bibinfo {year}
  {2022}{\natexlab{b}})}\BibitemShut {NoStop}%
\bibitem [{\citenamefont {Foster}\ \emph
  {et~al.}(2025{\natexlab{a}})\citenamefont {Foster} \emph
  {et~al.}}]{Foster:2025csl}%
  \BibitemOpen
  \bibfield  {author} {\bibinfo {author} {\bibfnamefont {J.~W.}\ \bibnamefont
  {Foster}} \emph {et~al.},\ }\href@noop {} {\bibfield  {journal} {\bibinfo
  {journal} {submitted to Phys. Rev. D}\ } (\bibinfo {year}
  {2025}{\natexlab{a}})},\ \Eprint {https://arxiv.org/abs/2504.16988}
  {arXiv:2504.16988 [gr-qc]} \BibitemShut {NoStop}%
\bibitem [{\citenamefont {Foster}\ \emph
  {et~al.}(2025{\natexlab{b}})\citenamefont {Foster} \emph
  {et~al.}}]{Foster:2025nzf}%
  \BibitemOpen
  \bibfield  {author} {\bibinfo {author} {\bibfnamefont {J.~W.}\ \bibnamefont
  {Foster}} \emph {et~al.},\ }\href@noop {} {\bibfield  {journal} {\bibinfo
  {journal} {submitted to Phys. Rev. Letters}\ } (\bibinfo {year}
  {2025}{\natexlab{b}})},\ \Eprint {https://arxiv.org/abs/2504.15334}
  {arXiv:2504.15334 [astro-ph.CO]} \BibitemShut {NoStop}%
\bibitem [{\citenamefont {{Misner}}\ \emph {et~al.}(1973)\citenamefont
  {{Misner}}, \citenamefont {{Thorne}},\ and\ \citenamefont
  {{Wheeler}}}]{misner:1973fk}%
  \BibitemOpen
  \bibfield  {author} {\bibinfo {author} {\bibfnamefont {C.~W.}\ \bibnamefont
  {{Misner}}}, \bibinfo {author} {\bibfnamefont {K.~S.}\ \bibnamefont
  {{Thorne}}},\ and\ \bibinfo {author} {\bibfnamefont {J.~A.}\ \bibnamefont
  {{Wheeler}}},\ }\href@noop {} {\emph {\bibinfo {title} {San Francisco:
  W.H.~Freeman and Co., 1973}}},\ edited by\ \bibinfo {editor} {\bibnamefont
  {{Misner, C.~W., Thorne K.~S. \& Wheeler J.~A.}}},\ {Physics Series}\
  (\bibinfo  {publisher} {{W. H. Freeman}},\ \bibinfo {year}
  {1973})\BibitemShut {NoStop}%
\bibitem [{\citenamefont {Thrane}\ and\ \citenamefont
  {Romano}(2013)}]{Thrane:2013oya}%
  \BibitemOpen
  \bibfield  {author} {\bibinfo {author} {\bibfnamefont {E.}~\bibnamefont
  {Thrane}}\ and\ \bibinfo {author} {\bibfnamefont {J.~D.}\ \bibnamefont
  {Romano}},\ }\href {https://doi.org/10.1103/PhysRevD.88.124032} {\bibfield
  {journal} {\bibinfo  {journal} {Phys. Rev. D}\ }\textbf {\bibinfo {volume}
  {88}},\ \bibinfo {pages} {124032} (\bibinfo {year} {2013})},\ \Eprint
  {https://arxiv.org/abs/1310.5300} {arXiv:1310.5300 [astro-ph.IM]}
  \BibitemShut {NoStop}%
\bibitem [{\citenamefont {Ellis}\ \emph {et~al.}(2023)\citenamefont {Ellis},
  \citenamefont {Fairbairn}, \citenamefont {H{\"u}tsi}, \citenamefont {Raidal},
  \citenamefont {Urrutia}, \citenamefont {Vaskonen},\ and\ \citenamefont
  {Veerm{\"a}e}}]{Ellis:2023owy}%
  \BibitemOpen
  \bibfield  {author} {\bibinfo {author} {\bibfnamefont {J.}~\bibnamefont
  {Ellis}}, \bibinfo {author} {\bibfnamefont {M.}~\bibnamefont {Fairbairn}},
  \bibinfo {author} {\bibfnamefont {G.}~\bibnamefont {H{\"u}tsi}}, \bibinfo
  {author} {\bibfnamefont {M.}~\bibnamefont {Raidal}}, \bibinfo {author}
  {\bibfnamefont {J.}~\bibnamefont {Urrutia}}, \bibinfo {author} {\bibfnamefont
  {V.}~\bibnamefont {Vaskonen}},\ and\ \bibinfo {author} {\bibfnamefont
  {H.}~\bibnamefont {Veerm{\"a}e}},\ }\href
  {https://doi.org/10.1051/0004-6361/202346268} {\bibfield  {journal} {\bibinfo
   {journal} {Astron. Astrophys.}\ }\textbf {\bibinfo {volume} {676}},\
  \bibinfo {pages} {A38} (\bibinfo {year} {2023})},\ \Eprint
  {https://arxiv.org/abs/2301.13854} {arXiv:2301.13854 [astro-ph.CO]}
  \BibitemShut {NoStop}%
\bibitem [{\citenamefont {Caprini}\ \emph {et~al.}(2016)\citenamefont {Caprini}
  \emph {et~al.}}]{Caprini:2015zlo}%
  \BibitemOpen
  \bibfield  {author} {\bibinfo {author} {\bibfnamefont {C.}~\bibnamefont
  {Caprini}} \emph {et~al.},\ }\href
  {https://doi.org/10.1088/1475-7516/2016/04/001} {\bibfield  {journal}
  {\bibinfo  {journal} {JCAP}\ }\textbf {\bibinfo {volume} {04}},\ \bibinfo
  {pages} {001}},\ \Eprint {https://arxiv.org/abs/1512.06239} {arXiv:1512.06239
  [astro-ph.CO]} \BibitemShut {NoStop}%
\bibitem [{\citenamefont {Auclair}\ \emph {et~al.}(2020)\citenamefont {Auclair}
  \emph {et~al.}}]{Auclair:2019wcv}%
  \BibitemOpen
  \bibfield  {author} {\bibinfo {author} {\bibfnamefont {P.}~\bibnamefont
  {Auclair}} \emph {et~al.},\ }\href
  {https://doi.org/10.1088/1475-7516/2020/04/034} {\bibfield  {journal}
  {\bibinfo  {journal} {JCAP}\ }\textbf {\bibinfo {volume} {04}},\ \bibinfo
  {pages} {034}},\ \Eprint {https://arxiv.org/abs/1909.00819} {arXiv:1909.00819
  [astro-ph.CO]} \BibitemShut {NoStop}%
\bibitem [{\citenamefont {Strugarek}\ \emph {et~al.}(2021)\citenamefont
  {Strugarek}, \citenamefont {So{\'s}nica}, \citenamefont {Arnold},
  \citenamefont {J{\"a}ggi}, \citenamefont {Zajdel},\ and\ \citenamefont
  {Bury}}]{Strugarek2026arcs}%
  \BibitemOpen
  \bibfield  {author} {\bibinfo {author} {\bibfnamefont {D.}~\bibnamefont
  {Strugarek}}, \bibinfo {author} {\bibfnamefont {K.}~\bibnamefont
  {So{\'s}nica}}, \bibinfo {author} {\bibfnamefont {D.}~\bibnamefont {Arnold}},
  \bibinfo {author} {\bibfnamefont {A.}~\bibnamefont {J{\"a}ggi}}, \bibinfo
  {author} {\bibfnamefont {R.}~\bibnamefont {Zajdel}},\ and\ \bibinfo {author}
  {\bibfnamefont {G.}~\bibnamefont {Bury}},\ }\href
  {https://doi.org/10.1186/s40623-021-01397-1} {\bibfield  {journal} {\bibinfo
  {journal} {Earth, Planets and Space}\ }\textbf {\bibinfo {volume} {73}},\
  \bibinfo {pages} {87} (\bibinfo {year} {2021})}\BibitemShut {NoStop}%
\bibitem [{\citenamefont {{Gourgoulhon}}\ \emph {et~al.}(2019)\citenamefont
  {{Gourgoulhon}}, \citenamefont {{Le Tiec}}, \citenamefont {{Vincent}},\ and\
  \citenamefont {{Warburton}}}]{gourgoulhon:2019aa}%
  \BibitemOpen
  \bibfield  {author} {\bibinfo {author} {\bibfnamefont {E.}~\bibnamefont
  {{Gourgoulhon}}}, \bibinfo {author} {\bibfnamefont {A.}~\bibnamefont {{Le
  Tiec}}}, \bibinfo {author} {\bibfnamefont {F.~H.}\ \bibnamefont
  {{Vincent}}},\ and\ \bibinfo {author} {\bibfnamefont {N.}~\bibnamefont
  {{Warburton}}},\ }\href@noop {} {\bibfield  {journal} {\bibinfo  {journal}
  {\aap}\ }\textbf {\bibinfo {volume} {627}},\ \bibinfo {pages} {A92} (\bibinfo
  {year} {2019})}\BibitemShut {NoStop}%
\bibitem [{\citenamefont {Caprini}\ and\ \citenamefont
  {Figueroa}(2018)}]{Caprini:2018mtu}%
  \BibitemOpen
  \bibfield  {author} {\bibinfo {author} {\bibfnamefont {C.}~\bibnamefont
  {Caprini}}\ and\ \bibinfo {author} {\bibfnamefont {D.~G.}\ \bibnamefont
  {Figueroa}},\ }\href {https://doi.org/10.1088/1361-6382/aac608} {\bibfield
  {journal} {\bibinfo  {journal} {Class. Quant. Grav.}\ }\textbf {\bibinfo
  {volume} {35}},\ \bibinfo {pages} {163001} (\bibinfo {year}
  {2018})}\BibitemShut {NoStop}%
\bibitem [{\citenamefont {Afzal}\ \emph {et~al.}(2023)\citenamefont {Afzal}
  \emph {et~al.}}]{NANOGrav:2023hvm}%
  \BibitemOpen
  \bibfield  {author} {\bibinfo {author} {\bibfnamefont {A.}~\bibnamefont
  {Afzal}} \emph {et~al.} (\bibinfo {collaboration} {NANOGrav}),\ }\href
  {https://doi.org/10.3847/2041-8213/acdc91} {\bibfield  {journal} {\bibinfo
  {journal} {Astrophys. J. Lett.}\ }\textbf {\bibinfo {volume} {951}},\
  \bibinfo {pages} {L11} (\bibinfo {year} {2023})},\ \bibinfo {note} {[Erratum:
  Astrophys.J.Lett. 971, L27 (2024), Erratum: Astrophys.J. 971, L27 (2024)]},\
  \Eprint {https://arxiv.org/abs/2306.16219} {arXiv:2306.16219 [astro-ph.HE]}
  \BibitemShut {NoStop}%
\bibitem [{EuropeanStrategy()}]{EUreport}%
  \BibitemOpen
  EuropeanStrategy,\ \href@noop {} {\bibinfo {title} {{European Strategy for
  Particle Physics }}},\ \bibinfo {howpublished}
  {\url{https://europeanstrategy.cern/}}\BibitemShut {NoStop}%
\bibitem [{APPECRoadmap()}]{APPEC}%
  \BibitemOpen
  APPECRoadmap,\ \href@noop {} {\bibinfo {title} {{European Astroparticle
  Physics Strategy Surveys}}},\ \bibinfo {howpublished}
  {\url{https://www.appec.org/roadmap/}}\BibitemShut {NoStop}%
\bibitem [{\citenamefont {Brito}\ \emph {et~al.}(2015)\citenamefont {Brito},
  \citenamefont {Cardoso},\ and\ \citenamefont {Pani}}]{Brito:2015oca}%
  \BibitemOpen
  \bibfield  {author} {\bibinfo {author} {\bibfnamefont {R.}~\bibnamefont
  {Brito}}, \bibinfo {author} {\bibfnamefont {V.}~\bibnamefont {Cardoso}},\
  and\ \bibinfo {author} {\bibfnamefont {P.}~\bibnamefont {Pani}},\ }\href
  {https://doi.org/10.1007/978-3-319-19000-6} {\bibfield  {journal} {\bibinfo
  {journal} {Lect. Notes Phys.}\ }\textbf {\bibinfo {volume} {906}},\ \bibinfo
  {pages} {pp.1} (\bibinfo {year} {2015})},\ \Eprint
  {https://arxiv.org/abs/1501.06570} {arXiv:1501.06570 [gr-qc]} \BibitemShut
  {NoStop}%
\bibitem [{\citenamefont {Blas}\ \emph {et~al.}(2026)\citenamefont {Blas},
  \citenamefont {Foster}, \citenamefont {Gouttenoire}, \citenamefont {Iovino},
  \citenamefont {Musco}, \citenamefont {Trifinopoulos},\ and\ \citenamefont
  {Vanvlasselaer}}]{Blas:2026xws}%
  \BibitemOpen
  \bibfield  {author} {\bibinfo {author} {\bibfnamefont {D.}~\bibnamefont
  {Blas}}, \bibinfo {author} {\bibfnamefont {J.~W.}\ \bibnamefont {Foster}},
  \bibinfo {author} {\bibfnamefont {Y.}~\bibnamefont {Gouttenoire}}, \bibinfo
  {author} {\bibfnamefont {A.~J.}\ \bibnamefont {Iovino}}, \bibinfo {author}
  {\bibfnamefont {I.}~\bibnamefont {Musco}}, \bibinfo {author} {\bibfnamefont
  {S.}~\bibnamefont {Trifinopoulos}},\ and\ \bibinfo {author} {\bibfnamefont
  {M.}~\bibnamefont {Vanvlasselaer}},\ }\href
  {https://doi.org/10.1016/j.physletb.2026.140635} {\bibfield  {journal}
  {\bibinfo  {journal} {Phys. Lett. B}\ }\textbf {\bibinfo {volume} {879}},\
  \bibinfo {pages} {140635} (\bibinfo {year} {2026})},\ \Eprint
  {https://arxiv.org/abs/2602.12252} {arXiv:2602.12252 [astro-ph.CO]}
  \BibitemShut {NoStop}%
\bibitem [{\citenamefont {Carr}\ \emph {et~al.}(2026)\citenamefont {Carr},
  \citenamefont {Iovino}, \citenamefont {Perna}, \citenamefont {Vaskonen},\
  and\ \citenamefont {Veerm{\"a}e}}]{Carr:2026hot}%
  \BibitemOpen
  \bibfield  {author} {\bibinfo {author} {\bibfnamefont {B.}~\bibnamefont
  {Carr}}, \bibinfo {author} {\bibfnamefont {A.~J.}\ \bibnamefont {Iovino}},
  \bibinfo {author} {\bibfnamefont {G.}~\bibnamefont {Perna}}, \bibinfo
  {author} {\bibfnamefont {V.}~\bibnamefont {Vaskonen}},\ and\ \bibinfo
  {author} {\bibfnamefont {H.}~\bibnamefont {Veerm{\"a}e}},\ }\href
  {https://doi.org/10.1007/s40766-026-00080-z} {\bibfield  {journal} {\bibinfo
  {journal} {Riv. Nuovo Cim.}\ }\textbf {\bibinfo {volume} {49}},\ \bibinfo
  {pages} {225} (\bibinfo {year} {2026})},\ \Eprint
  {https://arxiv.org/abs/2601.06024} {arXiv:2601.06024 [astro-ph.CO]}
  \BibitemShut {NoStop}%
\bibitem [{\citenamefont {Niikura}\ \emph {et~al.}(2019)\citenamefont
  {Niikura}, \citenamefont {Takada}, \citenamefont {Yokoyama}, \citenamefont
  {Sumi},\ and\ \citenamefont {Masaki}}]{Niikura:2019kqi}%
  \BibitemOpen
  \bibfield  {author} {\bibinfo {author} {\bibfnamefont {H.}~\bibnamefont
  {Niikura}}, \bibinfo {author} {\bibfnamefont {M.}~\bibnamefont {Takada}},
  \bibinfo {author} {\bibfnamefont {S.}~\bibnamefont {Yokoyama}}, \bibinfo
  {author} {\bibfnamefont {T.}~\bibnamefont {Sumi}},\ and\ \bibinfo {author}
  {\bibfnamefont {S.}~\bibnamefont {Masaki}},\ }\href
  {https://doi.org/10.1103/PhysRevD.99.083503} {\bibfield  {journal} {\bibinfo
  {journal} {Phys. Rev. D}\ }\textbf {\bibinfo {volume} {99}},\ \bibinfo
  {pages} {083503} (\bibinfo {year} {2019})},\ \Eprint
  {https://arxiv.org/abs/1901.07120} {arXiv:1901.07120 [astro-ph.CO]}
  \BibitemShut {NoStop}%
\bibitem [{\citenamefont {Sugiyama}\ \emph {et~al.}(2026)\citenamefont
  {Sugiyama}, \citenamefont {Takada}, \citenamefont {Yasuda},\ and\
  \citenamefont {Tominaga}}]{Sugiyama:2026kpv}%
  \BibitemOpen
  \bibfield  {author} {\bibinfo {author} {\bibfnamefont {S.}~\bibnamefont
  {Sugiyama}}, \bibinfo {author} {\bibfnamefont {M.}~\bibnamefont {Takada}},
  \bibinfo {author} {\bibfnamefont {N.}~\bibnamefont {Yasuda}},\ and\ \bibinfo
  {author} {\bibfnamefont {N.}~\bibnamefont {Tominaga}},\ }\href@noop {}
  {\bibfield  {journal} {\bibinfo  {journal} {arXiv}\ } (\bibinfo {year}
  {2026})},\ \Eprint {https://arxiv.org/abs/2602.05840} {arXiv:2602.05840
  [astro-ph.CO]} \BibitemShut {NoStop}%
\bibitem [{\citenamefont {{Moody}}\ and\ \citenamefont
  {{Wilczek}}(1984)}]{moody:1984tq}%
  \BibitemOpen
  \bibfield  {author} {\bibinfo {author} {\bibfnamefont {J.~E.}\ \bibnamefont
  {{Moody}}}\ and\ \bibinfo {author} {\bibfnamefont {F.}~\bibnamefont
  {{Wilczek}}},\ }\href@noop {} {\bibfield  {journal} {\bibinfo  {journal}
  {\prd}\ }\textbf {\bibinfo {volume} {30}},\ \bibinfo {pages} {130} (\bibinfo
  {year} {1984})}\BibitemShut {NoStop}%
\bibitem [{\citenamefont {{Fischbach}}\ and\ \citenamefont
  {{Talmadge}}(1999)}]{fischbach:1999ly}%
  \BibitemOpen
  \bibfield  {author} {\bibinfo {author} {\bibfnamefont {E.}~\bibnamefont
  {{Fischbach}}}\ and\ \bibinfo {author} {\bibfnamefont {C.~L.}\ \bibnamefont
  {{Talmadge}}},\ }\href@noop {} {\emph {\bibinfo {title} {{The Search for
  Non-Newtonian Gravity}}}},\ edited by\ \bibinfo {editor} {\bibnamefont
  {{Fischbach, E.~\& Talmadge, C.~L.}}},\ {Aip-Press Series}\ (\bibinfo
  {publisher} {{Springer}},\ \bibinfo {year} {1999})\BibitemShut {NoStop}%
\bibitem [{\citenamefont {{Lucchesi}}\ and\ \citenamefont
  {{Peron}}(2010)}]{2010PhRvL.105w1103L}%
  \BibitemOpen
  \bibfield  {author} {\bibinfo {author} {\bibfnamefont {D.~M.}\ \bibnamefont
  {{Lucchesi}}}\ and\ \bibinfo {author} {\bibfnamefont {R.}~\bibnamefont
  {{Peron}}},\ }\href {https://doi.org/10.1103/PhysRevLett.105.231103}
  {\bibfield  {journal} {\bibinfo  {journal} {Phys. Rev. Lett.}\ }\textbf
  {\bibinfo {volume} {105}},\ \bibinfo {eid} {231103} (\bibinfo {year}
  {2010})},\ \Eprint {https://arxiv.org/abs/1106.2905} {arXiv:1106.2905
  [gr-qc]} \BibitemShut {NoStop}%
\bibitem [{\citenamefont {Lucchesi}\ and\ \citenamefont
  {Peron}(2014)}]{Lucchesi:2014uza}%
  \BibitemOpen
  \bibfield  {author} {\bibinfo {author} {\bibfnamefont {D.~M.}\ \bibnamefont
  {Lucchesi}}\ and\ \bibinfo {author} {\bibfnamefont {R.}~\bibnamefont
  {Peron}},\ }\href {https://doi.org/10.1103/PhysRevD.89.082002} {\bibfield
  {journal} {\bibinfo  {journal} {Phys. Rev. D}\ }\textbf {\bibinfo {volume}
  {89}},\ \bibinfo {pages} {082002} (\bibinfo {year} {2014})}\BibitemShut
  {NoStop}%
\bibitem [{\citenamefont {{Bailey}}\ and\ \citenamefont
  {{Kosteleck{\'y}}}(2006)}]{bailey:2006uq}%
  \BibitemOpen
  \bibfield  {author} {\bibinfo {author} {\bibfnamefont {Q.~G.}\ \bibnamefont
  {{Bailey}}}\ and\ \bibinfo {author} {\bibfnamefont {V.~A.}\ \bibnamefont
  {{Kosteleck{\'y}}}},\ }\href@noop {} {\bibfield  {journal} {\bibinfo
  {journal} {\prd}\ }\textbf {\bibinfo {volume} {74}},\ \bibinfo {pages}
  {045001} (\bibinfo {year} {2006})}\BibitemShut {NoStop}%
\bibitem [{\citenamefont {Lucchesi}\ \emph {et~al.}(2026)\citenamefont
  {Lucchesi}, \citenamefont {Visco}, \citenamefont {Peron}, \citenamefont
  {Rodriguez}, \citenamefont {Pucacco}, \citenamefont {Anselmo}, \citenamefont
  {Bassan}, \citenamefont {Appleby}, \citenamefont {Cinelli}, \citenamefont
  {Marco}, \citenamefont {Lucente}, \citenamefont {Magnafico}, \citenamefont
  {Pardini},\ and\ \citenamefont {Sapio}}]{lucchesi:2026aa}%
  \BibitemOpen
  \bibfield  {author} {\bibinfo {author} {\bibfnamefont {D.}~\bibnamefont
  {Lucchesi}}, \bibinfo {author} {\bibfnamefont {M.}~\bibnamefont {Visco}},
  \bibinfo {author} {\bibfnamefont {R.}~\bibnamefont {Peron}}, \bibinfo
  {author} {\bibfnamefont {J.~C.}\ \bibnamefont {Rodriguez}}, \bibinfo {author}
  {\bibfnamefont {G.}~\bibnamefont {Pucacco}}, \bibinfo {author} {\bibfnamefont
  {L.}~\bibnamefont {Anselmo}}, \bibinfo {author} {\bibfnamefont
  {M.}~\bibnamefont {Bassan}}, \bibinfo {author} {\bibfnamefont
  {G.}~\bibnamefont {Appleby}}, \bibinfo {author} {\bibfnamefont
  {M.}~\bibnamefont {Cinelli}}, \bibinfo {author} {\bibfnamefont {A.~D.}\
  \bibnamefont {Marco}}, \bibinfo {author} {\bibfnamefont {M.}~\bibnamefont
  {Lucente}}, \bibinfo {author} {\bibfnamefont {C.}~\bibnamefont {Magnafico}},
  \bibinfo {author} {\bibfnamefont {C.}~\bibnamefont {Pardini}},\ and\ \bibinfo
  {author} {\bibfnamefont {F.}~\bibnamefont {Sapio}},\ }\href
  {https://doi.org/10.1103/hj6p-bfyr} {\bibfield  {journal} {\bibinfo
  {journal} {Phys. Rev. D}\ ,\ } (\bibinfo {year} {2026})}\BibitemShut
  {NoStop}%
\bibitem [{\citenamefont {Lucchesi}\ \emph {et~al.}(2020)\citenamefont
  {Lucchesi}, \citenamefont {Anselmo}, \citenamefont {Bassan}, \citenamefont
  {Magnafico}, \citenamefont {Pardini}, \citenamefont {Peron}, \citenamefont
  {Pucacco},\ and\ \citenamefont {Visco}}]{Lucchesi2020}%
  \BibitemOpen
  \bibfield  {author} {\bibinfo {author} {\bibfnamefont {D.~M.}\ \bibnamefont
  {Lucchesi}}, \bibinfo {author} {\bibfnamefont {L.}~\bibnamefont {Anselmo}},
  \bibinfo {author} {\bibfnamefont {M.}~\bibnamefont {Bassan}}, \bibinfo
  {author} {\bibfnamefont {C.}~\bibnamefont {Magnafico}}, \bibinfo {author}
  {\bibfnamefont {C.}~\bibnamefont {Pardini}}, \bibinfo {author} {\bibfnamefont
  {R.}~\bibnamefont {Peron}}, \bibinfo {author} {\bibfnamefont
  {G.}~\bibnamefont {Pucacco}},\ and\ \bibinfo {author} {\bibfnamefont
  {M.}~\bibnamefont {Visco}},\ }\href {https://doi.org/10.3390/universe6090139}
  {\bibfield  {journal} {\bibinfo  {journal} {Universe}\ }\textbf {\bibinfo
  {volume} {6}},\ \bibinfo {pages} {139} (\bibinfo {year} {2020})}\BibitemShut
  {NoStop}%
\bibitem [{\citenamefont {Bauer}\ and\ \citenamefont
  {Rostagni}(2024)}]{Bauer:2023czj}%
  \BibitemOpen
  \bibfield  {author} {\bibinfo {author} {\bibfnamefont {M.}~\bibnamefont
  {Bauer}}\ and\ \bibinfo {author} {\bibfnamefont {G.}~\bibnamefont
  {Rostagni}},\ }\href {https://doi.org/10.1103/PhysRevLett.132.101802}
  {\bibfield  {journal} {\bibinfo  {journal} {Phys. Rev. Lett.}\ }\textbf
  {\bibinfo {volume} {132}},\ \bibinfo {pages} {101802} (\bibinfo {year}
  {2024})},\ \Eprint {https://arxiv.org/abs/2307.09516} {arXiv:2307.09516
  [hep-ph]} \BibitemShut {NoStop}%
\bibitem [{\citenamefont {Grossman}\ \emph {et~al.}(2025)\citenamefont
  {Grossman}, \citenamefont {Yu},\ and\ \citenamefont
  {Zhou}}]{Grossman:2025cov}%
  \BibitemOpen
  \bibfield  {author} {\bibinfo {author} {\bibfnamefont {Y.}~\bibnamefont
  {Grossman}}, \bibinfo {author} {\bibfnamefont {B.}~\bibnamefont {Yu}},\ and\
  \bibinfo {author} {\bibfnamefont {S.}~\bibnamefont {Zhou}},\ }\href@noop {}
  {\  (\bibinfo {year} {2025})},\ \Eprint {https://arxiv.org/abs/2504.00104}
  {arXiv:2504.00104 [hep-ph]} \BibitemShut {NoStop}%
\bibitem [{\citenamefont {O'Connell}\ \emph {et~al.}(2007)\citenamefont
  {O'Connell}, \citenamefont {Ramsey-Musolf},\ and\ \citenamefont
  {Wise}}]{OConnell:2006rsp}%
  \BibitemOpen
  \bibfield  {author} {\bibinfo {author} {\bibfnamefont {D.}~\bibnamefont
  {O'Connell}}, \bibinfo {author} {\bibfnamefont {M.~J.}\ \bibnamefont
  {Ramsey-Musolf}},\ and\ \bibinfo {author} {\bibfnamefont {M.~B.}\
  \bibnamefont {Wise}},\ }\href {https://doi.org/10.1103/PhysRevD.75.037701}
  {\bibfield  {journal} {\bibinfo  {journal} {Phys. Rev. D}\ }\textbf {\bibinfo
  {volume} {75}},\ \bibinfo {pages} {037701} (\bibinfo {year} {2007})},\
  \Eprint {https://arxiv.org/abs/hep-ph/0611014} {arXiv:hep-ph/0611014}
  \BibitemShut {NoStop}%
\bibitem [{\citenamefont {Patt}\ and\ \citenamefont
  {Wilczek}(2006)}]{Patt:2006fw}%
  \BibitemOpen
  \bibfield  {author} {\bibinfo {author} {\bibfnamefont {B.}~\bibnamefont
  {Patt}}\ and\ \bibinfo {author} {\bibfnamefont {F.}~\bibnamefont {Wilczek}},\
  }\href@noop {} {\  (\bibinfo {year} {2006})},\ \Eprint
  {https://arxiv.org/abs/hep-ph/0605188} {arXiv:hep-ph/0605188} \BibitemShut
  {NoStop}%
\bibitem [{\citenamefont {Burrage}\ \emph
  {et~al.}(2026{\natexlab{a}})\citenamefont {Burrage}, \citenamefont
  {Macdonald}, \citenamefont {Ross}, \citenamefont {Rybka},\ and\ \citenamefont
  {Todarello}}]{Burrage:2025grx}%
  \BibitemOpen
  \bibfield  {author} {\bibinfo {author} {\bibfnamefont {C.}~\bibnamefont
  {Burrage}}, \bibinfo {author} {\bibfnamefont {A.}~\bibnamefont {Macdonald}},
  \bibinfo {author} {\bibfnamefont {M.~P.}\ \bibnamefont {Ross}}, \bibinfo
  {author} {\bibfnamefont {G.}~\bibnamefont {Rybka}},\ and\ \bibinfo {author}
  {\bibfnamefont {E.}~\bibnamefont {Todarello}},\ }\href
  {https://doi.org/10.1103/bml1-j932} {\bibfield  {journal} {\bibinfo
  {journal} {Phys. Rev. D}\ }\textbf {\bibinfo {volume} {113}},\ \bibinfo
  {pages} {063505} (\bibinfo {year} {2026}{\natexlab{a}})},\ \Eprint
  {https://arxiv.org/abs/2507.16526} {arXiv:2507.16526 [hep-ph]} \BibitemShut
  {NoStop}%
\bibitem [{\citenamefont {Burrage}\ \emph
  {et~al.}(2026{\natexlab{b}})\citenamefont {Burrage}, \citenamefont
  {Macdonald},\ and\ \citenamefont {Todarello}}]{Burrage:2026loe}%
  \BibitemOpen
  \bibfield  {author} {\bibinfo {author} {\bibfnamefont {C.}~\bibnamefont
  {Burrage}}, \bibinfo {author} {\bibfnamefont {A.}~\bibnamefont {Macdonald}},\
  and\ \bibinfo {author} {\bibfnamefont {E.}~\bibnamefont {Todarello}},\
  }\href@noop {} {\bibinfo {title} {{Using the Pericentre Precession of LAGEOS
  II to Constrain Quadratically Coupled Ultralight Dark Matter}}} (\bibinfo
  {year} {2026}{\natexlab{b}}),\ \Eprint {https://arxiv.org/abs/2605.28248}
  {arXiv:2605.28248 [hep-ph]} \BibitemShut {NoStop}%
\bibitem [{\citenamefont {Hees}\ \emph {et~al.}(2018)\citenamefont {Hees},
  \citenamefont {Minazzoli}, \citenamefont {Savalle}, \citenamefont {Stadnik},\
  and\ \citenamefont {Wolf}}]{Hees:2018fpg}%
  \BibitemOpen
  \bibfield  {author} {\bibinfo {author} {\bibfnamefont {A.}~\bibnamefont
  {Hees}}, \bibinfo {author} {\bibfnamefont {O.}~\bibnamefont {Minazzoli}},
  \bibinfo {author} {\bibfnamefont {E.}~\bibnamefont {Savalle}}, \bibinfo
  {author} {\bibfnamefont {Y.~V.}\ \bibnamefont {Stadnik}},\ and\ \bibinfo
  {author} {\bibfnamefont {P.}~\bibnamefont {Wolf}},\ }\href
  {https://doi.org/10.1103/PhysRevD.98.064051} {\bibfield  {journal} {\bibinfo
  {journal} {Phys. Rev. D}\ }\textbf {\bibinfo {volume} {98}},\ \bibinfo
  {pages} {064051} (\bibinfo {year} {2018})},\ \Eprint
  {https://arxiv.org/abs/1807.04512} {arXiv:1807.04512 [gr-qc]} \BibitemShut
  {NoStop}%
\bibitem [{\citenamefont {Damour}\ and\ \citenamefont
  {Donoghue}(2010)}]{Damour:2010rp}%
  \BibitemOpen
  \bibfield  {author} {\bibinfo {author} {\bibfnamefont {T.}~\bibnamefont
  {Damour}}\ and\ \bibinfo {author} {\bibfnamefont {J.~F.}\ \bibnamefont
  {Donoghue}},\ }\href {https://doi.org/10.1103/PhysRevD.82.084033} {\bibfield
  {journal} {\bibinfo  {journal} {Phys. Rev. D}\ }\textbf {\bibinfo {volume}
  {82}},\ \bibinfo {pages} {084033} (\bibinfo {year} {2010})},\ \Eprint
  {https://arxiv.org/abs/1007.2792} {arXiv:1007.2792 [gr-qc]} \BibitemShut
  {NoStop}%
\bibitem [{\citenamefont {Delva}\ \emph {et~al.}(2023)\citenamefont {Delva},
  \citenamefont {Altamimi}, \citenamefont {Blazquez}, \citenamefont
  {Blossfeld}, \citenamefont {B{\"o}hm}, \citenamefont {Bonnefond},
  \citenamefont {Boy} \emph {et~al.}}]{Delva2023}%
  \BibitemOpen
  \bibfield  {author} {\bibinfo {author} {\bibfnamefont {P.}~\bibnamefont
  {Delva}}, \bibinfo {author} {\bibfnamefont {Z.}~\bibnamefont {Altamimi}},
  \bibinfo {author} {\bibfnamefont {A.}~\bibnamefont {Blazquez}}, \bibinfo
  {author} {\bibfnamefont {M.}~\bibnamefont {Blossfeld}}, \bibinfo {author}
  {\bibfnamefont {J.}~\bibnamefont {B{\"o}hm}}, \bibinfo {author}
  {\bibfnamefont {P.}~\bibnamefont {Bonnefond}}, \bibinfo {author}
  {\bibfnamefont {J.-P.}\ \bibnamefont {Boy}}, \emph {et~al.},\ }\href
  {https://doi.org/10.1186/s40623-022-01752-w} {\bibfield  {journal} {\bibinfo
  {journal} {Earth Planets Space}\ }\textbf {\bibinfo {volume} {75}},\ \bibinfo
  {pages} {5} (\bibinfo {year} {2023})}\BibitemShut {NoStop}%
\bibitem [{\citenamefont {Ries}\ \emph {et~al.}(1992)\citenamefont {Ries},
  \citenamefont {Eanes}, \citenamefont {Shum},\ and\ \citenamefont
  {Watkins}}]{Ries:1992}%
  \BibitemOpen
  \bibfield  {author} {\bibinfo {author} {\bibfnamefont {J.~C.}\ \bibnamefont
  {Ries}}, \bibinfo {author} {\bibfnamefont {R.~J.}\ \bibnamefont {Eanes}},
  \bibinfo {author} {\bibfnamefont {C.~K.}\ \bibnamefont {Shum}},\ and\
  \bibinfo {author} {\bibfnamefont {M.~M.}\ \bibnamefont {Watkins}},\
  }\href@noop {} {\bibfield  {journal} {\bibinfo  {journal} {Geophysical
  Research Letters}\ }\textbf {\bibinfo {volume} {19}},\ \bibinfo {pages} {529}
  (\bibinfo {year} {1992})}\BibitemShut {NoStop}%
\bibitem [{\citenamefont {Plag}\ and\ \citenamefont
  {Pearlman}(2009)}]{Plag2009}%
  \BibitemOpen
  \bibinfo {editor} {\bibfnamefont {H.-P.}\ \bibnamefont {Plag}}\ and\ \bibinfo
  {editor} {\bibfnamefont {M.}~\bibnamefont {Pearlman}},\ eds.,\ \href
  {https://doi.org/10.1007/978-3-642-02687-4} {\emph {\bibinfo {title} {Global
  Geodetic Observing System: Meeting the Requirements of a Global Society on a
  Changing Planet in 2020}}}\ (\bibinfo  {publisher} {Springer},\ \bibinfo
  {address} {Berlin, Heidelberg},\ \bibinfo {year} {2009})\ p.\ \bibinfo
  {pages} {332}\BibitemShut {NoStop}%
\bibitem [{\citenamefont {{Visco}}\ and\ \citenamefont
  {{Lucchesi}}(2016)}]{visco2016}%
  \BibitemOpen
  \bibfield  {author} {\bibinfo {author} {\bibfnamefont {M.}~\bibnamefont
  {{Visco}}}\ and\ \bibinfo {author} {\bibfnamefont {D.~M.}\ \bibnamefont
  {{Lucchesi}}},\ }\href {https://doi.org/10.1016/j.asr.2016.02.006} {\bibfield
   {journal} {\bibinfo  {journal} {Advances in Space Research}\ }\textbf
  {\bibinfo {volume} {57}},\ \bibinfo {pages} {1928} (\bibinfo {year}
  {2016})}\BibitemShut {NoStop}%
\bibitem [{\citenamefont {{Visco}}\ and\ \citenamefont
  {{Lucchesi}}(2018)}]{visco2018}%
  \BibitemOpen
  \bibfield  {author} {\bibinfo {author} {\bibfnamefont {M.}~\bibnamefont
  {{Visco}}}\ and\ \bibinfo {author} {\bibfnamefont {D.~M.}\ \bibnamefont
  {{Lucchesi}}},\ }\href {https://doi.org/10.1103/PhysRevD.98.044034}
  {\bibfield  {journal} {\bibinfo  {journal} {\prd}\ }\textbf {\bibinfo
  {volume} {98}},\ \bibinfo {eid} {044034} (\bibinfo {year}
  {2018})}\BibitemShut {NoStop}%
\bibitem [{\citenamefont {Geisser}(2023)}]{Geisser2023}%
  \BibitemOpen
  \bibfield  {author} {\bibinfo {author} {\bibfnamefont {L.}~\bibnamefont
  {Geisser}},\ }\emph {\bibinfo {title} {Generation and Analysis of Satellite
  Laser Ranging Normal Points for Geodetic Parameter Estimation}},\ \href
  {https://boristheses.unibe.ch/4855/1/23geisser_l.pdf} {Ph.D. thesis},\
  \bibinfo  {school} {Universit{\"a}t Bern} (\bibinfo {year}
  {2023})\BibitemShut {NoStop}%
\bibitem [{\citenamefont {{So{\'s}nica}}\ \emph {et~al.}(2025)\citenamefont
  {{So{\'s}nica}}, \citenamefont {{Ga{\l}dyn}}, \citenamefont {{Zajdel}},
  \citenamefont {{Strugarek}}, \citenamefont {{Najder}}, \citenamefont
  {{Nowak}}, \citenamefont {{Miko{\'s}}}, \citenamefont {{Kur}}, \citenamefont
  {{Bosy}},\ and\ \citenamefont {{Bury}}}]{Sosnica2025a}%
  \BibitemOpen
  \bibfield  {author} {\bibinfo {author} {\bibfnamefont {K.}~\bibnamefont
  {{So{\'s}nica}}}, \bibinfo {author} {\bibfnamefont {F.}~\bibnamefont
  {{Ga{\l}dyn}}}, \bibinfo {author} {\bibfnamefont {R.}~\bibnamefont
  {{Zajdel}}}, \bibinfo {author} {\bibfnamefont {D.}~\bibnamefont
  {{Strugarek}}}, \bibinfo {author} {\bibfnamefont {J.}~\bibnamefont
  {{Najder}}}, \bibinfo {author} {\bibfnamefont {A.}~\bibnamefont {{Nowak}}},
  \bibinfo {author} {\bibfnamefont {M.}~\bibnamefont {{Miko{\'s}}}}, \bibinfo
  {author} {\bibfnamefont {T.}~\bibnamefont {{Kur}}}, \bibinfo {author}
  {\bibfnamefont {J.}~\bibnamefont {{Bosy}}},\ and\ \bibinfo {author}
  {\bibfnamefont {G.}~\bibnamefont {{Bury}}},\ }\href
  {https://doi.org/10.1007/s00190-024-01925-3} {\bibfield  {journal} {\bibinfo
  {journal} {Journal of Geodesy}\ }\textbf {\bibinfo {volume} {99}},\ \bibinfo
  {eid} {1} (\bibinfo {year} {2025})}\BibitemShut {NoStop}%
\end{thebibliography}%

\end{document}